\documentclass[iop, apj, twocolappendix, numberedappendix, appendixfloats]{emulateapj}
\usepackage{threeparttable}
\usepackage{multirow}
\usepackage{times}
\usepackage{enumitem}
\usepackage{url} 
\usepackage{color} 
\usepackage{amsmath,bm}

\shorttitle{The LAMOST and Gaia red clump sample}
\shortauthors{Y. Huang et al.}

\begin{document}

\title{Mapping the Galactic disk with the LAMOST and Gaia Red clump sample:\\ I:  precise distances, masses, ages and 3D velocities of $\sim$ 140\,000 red clump stars}


\author{Yang Huang\altaffilmark{1}}
\author{Ralph Sch{\"o}nrich\altaffilmark{2}}
\author{ Huawei Zhang\altaffilmark{3,4}}
\author{Yaqian Wu\altaffilmark{5}}
\author{ Bingqiu Chen\altaffilmark{1}}
\author{ Haifeng Wang\altaffilmark{1,9}}
\author{ Maosheng Xiang\altaffilmark{5,6}}
\author{Chun Wang\altaffilmark{3,9}}
\author{Haibo Yuan\altaffilmark{7}}
\author{Xinyi Li\altaffilmark{1}}
\author{Weixiang Sun\altaffilmark{1}}
\author{Ji Li\altaffilmark{8}}
\author{ Xiaowei Liu\altaffilmark{1}}
\altaffiltext{1}{South-Western Institute for Astronomy Research, Yunnan University, Kunming 650500, People's Republic of China; {\it yanghuang@ynu.edu.cn {\rm (YH)}; x.liu@ynu.edu.cn {\rm (XWL)}}}
\altaffiltext{2}{Mullard Space Science Laboratory, University College London, Holmbury St. Mary, Dorking, Surrey, RH5 6NT, UK; {\it r.schoenrich@ucl.ac.uk}}
\altaffiltext{3}{Department of Astronomy, Peking University, Beijing 100871, People's Republic of China}
\altaffiltext{4}{Kavli Institute for Astronomy and Astrophysics, Peking University, Beijing 100871, People's Republic of China}
\altaffiltext{5}{National Astronomical Observatories, Chinese Academy of Sciences, Beijing 100012, People's Republic of China}
\altaffiltext{6}{Max-Planck Institute for Astronomy, K{\"o}nigstuhl, D-69117, Heidelberg, Germany}
\altaffiltext{7}{Department of Astronomy, Beijing Normal University, Beijing 100875, People's Republic of China}
\altaffiltext{8}{Department of Space Science and Astronomy, Hebei Normal University, Shijiazhuang 050024, People's Republic of China}
\altaffiltext{9}{LAMOST Fellow}

\begin{abstract}
We present a sample of $\sim$\,140,000 primary red clump (RC) stars of spectral signal-to-noise ratios higher than 20 from the LAMOST Galactic spectroscopic surveys, selected based on their positions in the metallicity-dependent effective temperature--surface gravity and color--metallicity diagrams, supervised by high-quality $Kepler$ asteroseismology data. 
The stellar masses and ages of those stars are further determined from the LAMOST spectra, using the Kernel Principal Component Analysis method, trained with thousands of RCs in the  LAMOST-$Kepler$ fields with accurate asteroseismic mass measurements.  
The purity and completeness of our primary RC sample are generally higher than 80\,per cent.
For the mass and age, a variety of tests show typical uncertainties of 15 and 30\,per cent, respectively. 
Using over ten thousand primary RCs with accurate distance measurements from the parallaxes of Gaia\,DR2, we re-calibrate the $K_{\rm s}$ absolute magnitudes of primary RCs by, for the first time, considering both the metallicity and age dependencies. 
With the the new calibration, distances are derived for all the primary RCs, with a typical uncertainty of 5--10\,per cent,  even better than the values yielded by the Gaia parallax measurements for stars beyond 3--4\,kpc. 
The sample covers a significant volume of the Galactic disk of $4\,\leq\,R\,\leq\,16$\,kpc, $|Z|\,\leq\,5$\,kpc, and $-20\,\leq\,\phi\,\leq\,50^{\circ}$. 
Stellar atmospheric parameters, line-of-sight velocities and elemental abundances derived from the LAMOST spectra and proper motions of Gaia\,DR2 are also provided for the sample stars. 
Finally, the selection function of the sample is carefully evaluated in the color-magnitude plane for different sky areas. 
The sample is publicly available.
\end{abstract}\keywords{Galaxy: structure -- stars: distances -- stars: masses -- stars: ages}
%Core helium-burning primary red clump (RC) stars are considered to be standard candles since their absolute magnitudes are fairly independent on the stellar chemical composition and age.
%Given their large population in the Milky Way disk, they are ideal tracers to reveal the three-dimensional (3D) structure and its evolution of the Galactic disk.

\section{Introduction}
%What's RC and its importance
Primary red clump (RC) stars\footnote{In contrast, secondary RCs are helium burning (ignited non-degenerately) descendants of high mass (typically greater than 2\,$M_{\odot}$) stars and are non-standard candles} are metal-rich low-mass stars (typically smaller than $2 M_\odot$) of intermediate- to old-age in the core helium-burning phase (ignited degenerately).
They are widely used as distance indicators given their quite stable luminosities that are weakly dependent on chemical composition and age (e.g. Cannon 1970; Paczy{\'n}ski \& Stanek 1998).
By a careful calibration of the absolute magnitude of RC stars using the {\it Hipparcos} parallaxes (ESA 1997) of few hundred nearby RCs, precise distances to the Galactic center (Paczy{\'n}ski \& Stanek 1998), to the Local Group galaxies, e.g.  the Large Magellanic Cloud (LMC; Stanek, Zaritsky\,\& Harris 1998), the Small Magellanic Cloud (SMC; Subramanian \& Subramaniam 2012), and the Andromeda (M31; Stanek \& Garnavich 1998), have been determined.
 
 \begin{figure*}
\begin{center}
\includegraphics[scale=0.39,angle=0]{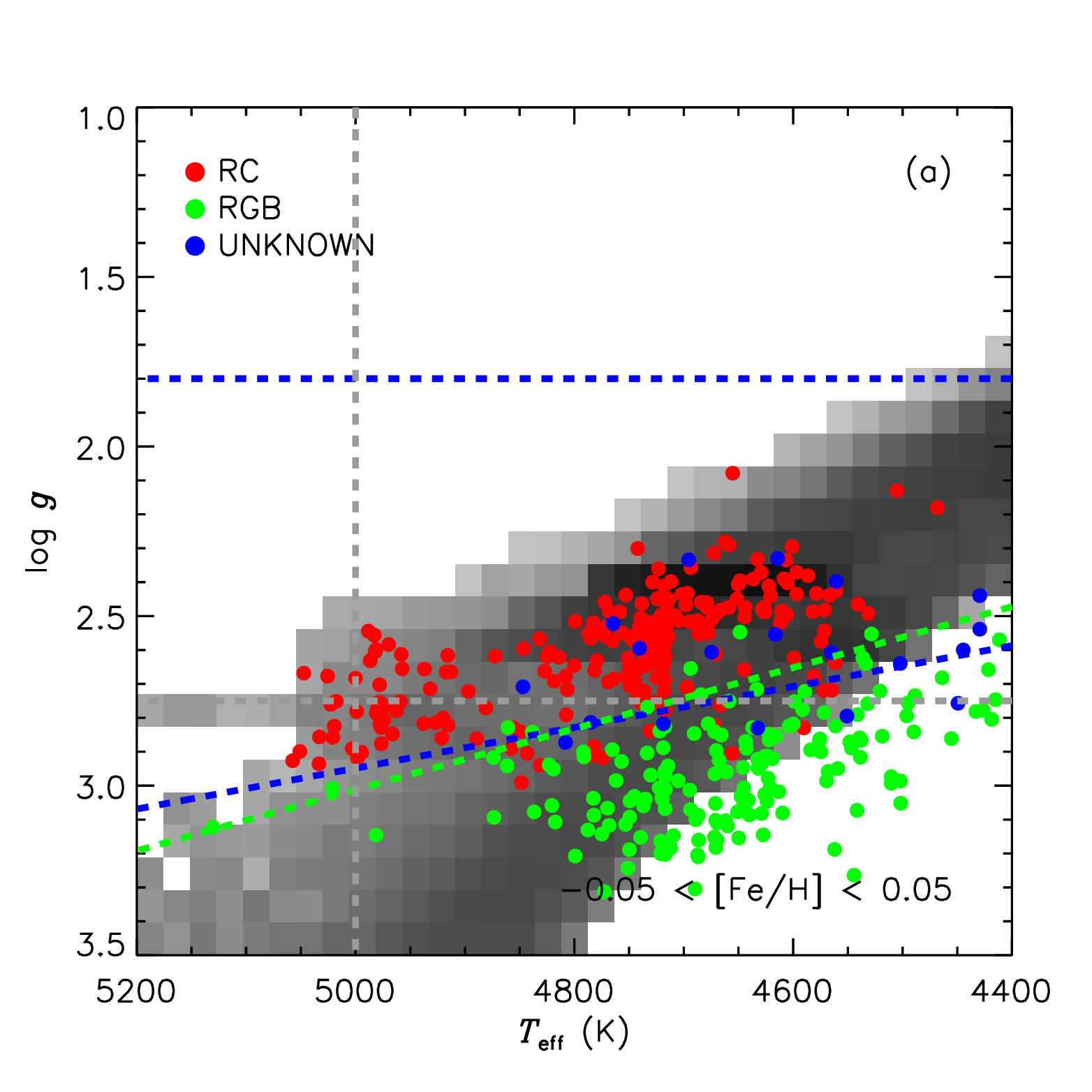}
\includegraphics[scale=0.40,angle=0]{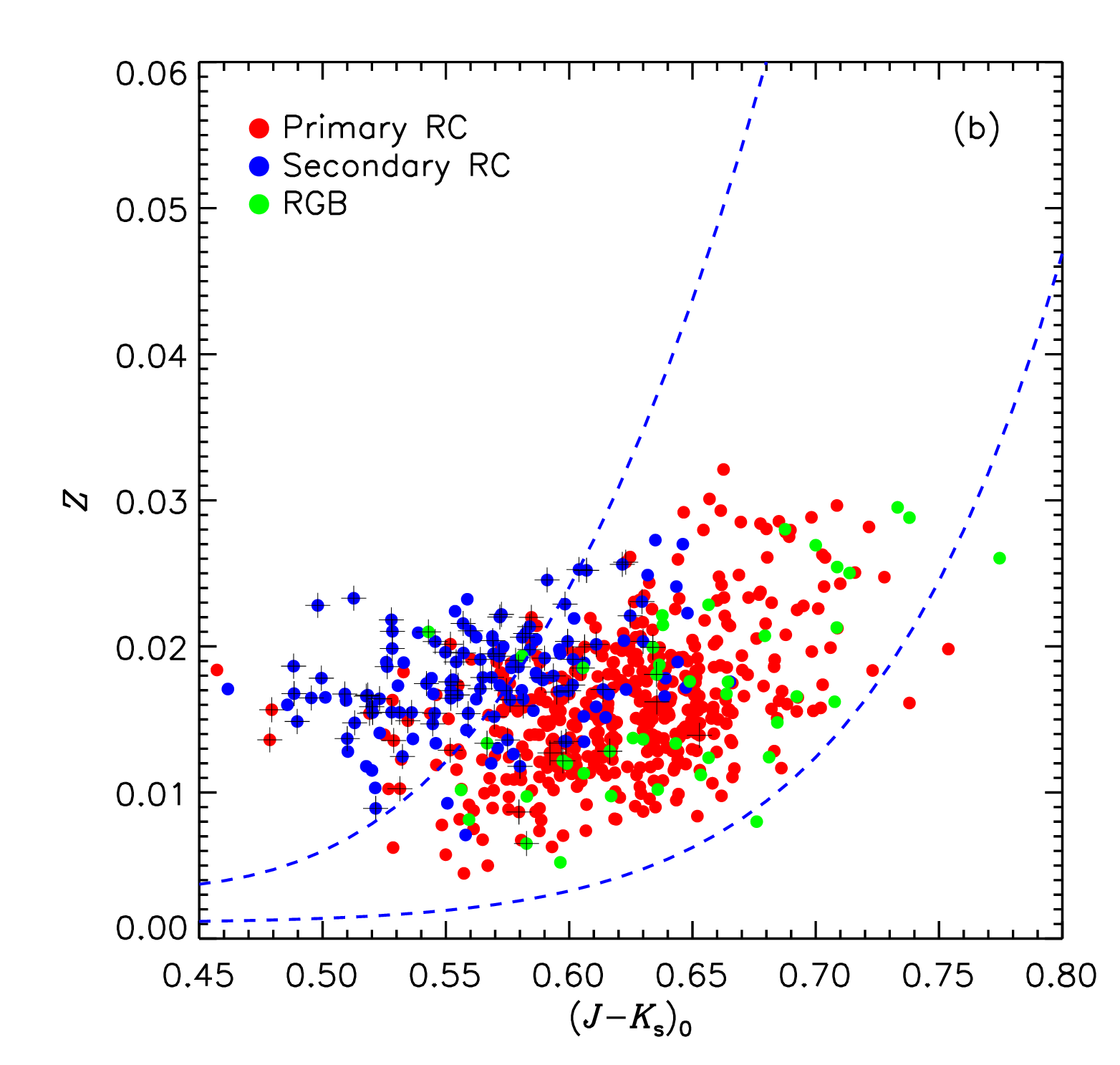}
\caption{{\it Left:} Distribution of LE sample stars in the $T_{\rm eff}$-log\,$g$ plane. 
The overplotted background (in grey scale) represents the distribution predicted by the PARSEC stellar evolution model (see text for details).
Values of $T_{\rm eff}$ and log\,$g$ are from the latest version of LSP3.
Red, green and blue dots represent, respectively, stars in the RC, RGB and unknown evolutionary stages as classified by Stello et al. (2013). 
The blue dashed lines denote the cuts, defined by Eq.\,1, to separate RCs from the less luminous RGBs.
The green dashed line represents the cut developed in H15. The slight difference between the cut in H15 and that in the current work is mainly due to the update of surface gravity measurements by Xiang et al. (2017a).
{The grey dashed lines represent the additional cuts on $T_{\rm eff}$ ($< 5000$\,K) and log\,$g$ ($< 2.75$) to further suppress the conamination of secondary RCs and RGBs.}
{\it Right:} Distribution of RC candidates from the LE sample in the color $(J-K_{\rm s})_0$-metallicity $Z$ plane, after passing the cuts defined by Eq.\,(1).
Red, blue and green dots denote the primary, secondary RCs and RGBs, respectively. 
Here primary and secondary RCs are defined by RCs with masses smaller than and greater than $2 M_{\odot}$, respectively.
The blue dashed lines represent the cuts, defined by Eqs.\,(3) and (4) to further eliminate contamination from the secondary RCs.
The black pluses mark stars of $T_{\rm eff} > 5000$\,K or log\,$g > 2.75$.}
\end{center}
\end{figure*}

%The selection of RC
As standard candles widely distributed across the entire Galactic disk, RCs are excellent tracers to explore the three-dimensional structure (e.g. Bovy et al. 2016), study the chemical and kinematic properties (e.g. Bovy et al. 2014, 2015; Nidever et al. 2014; Huang et al. 2015, 2016) and unravel the assemblage history of the Galactic disk.
However, it is a challenging task to identify the individual RCs amongst the numerous field stars.
In principle, the separation between RCs and other type stars, e.g., the red giant branch (RGB) stars can be carried out based on their large frequency separation $\Delta \nu$ of the acoustic modes and period spacing $\Delta P$ of the gravity modes measured from high-precision light curves, e.g. obtained with the {\it Kepler} and {\it CoRoT} satellites.
However, at the moment no more than a few ten thousand stars have $\Delta \nu$ and $\Delta P$ measurements in the {$\it Kepler$} and {$\it CoRoT$} fields.
More generally, with values of effective temperature $T_{\rm eff}$ and surface gravity log\,$g$ available from {the} large-scale spectroscopic surveys, such as RAVE and LAMOST for huge numbers of stars, a large number of RC candidates can be collected by selecting stars locating in a relatively ``small box" in the $T_{\rm eff}$ -- log\,$g$ diagram of $4800 \leq T_{\rm eff} \leq 5200$\,K and $2.0 \leq$\,log\,$g \leq 3.0$ (e.g. Siebert et al. 2011; Williams et al. 2013).
Unfortunately, RCs thus selected suffer from significant contamination, typically from 20 to 50 per cent (e.g. Williams et al. 2013; Wan et al. 2015), from RGB stars. 
By applying further cuts in the metallicity-dependent $T_{\rm eff}$ -- log\,$g$ plane and in the color $(J-K_{\rm s})_0$ -- metallicity $Z$ plane, Bovy et al. (2014) have successfully selected over ten thousand RC-like stars with a purity greater than 97 per cent from the APOGEE survey.
After calibration with the stellar evolution model and the high-quality asteroseismology data from {\it Kepler}, the {cut in the metallicity-dependent $T_{\rm eff}$ -- log\,$g$ plane} can separate the RC-like and RGB stars very well.
The {second cut in the color $(J-K_{\rm s})_0$ -- metallicity $Z$ plane} can further eliminate contamination from the secondary RCs.
The distances of the selected RCs are then estimated with a precision better than 5-10\,per cent.
By applying similar cuts to stars selected from the LAMOST Galactic spectroscopic surveys, Huang et al. (2015; hereafter H15) have obtained a clean RC sample of nearly hundred thousand stars, covering a significant volume of the Galactic disk.
More recently, using training set of stars selected from the {\it Kepler} asteroseismology data, Ting et al. (2018) have derived spectroscopy based estimates for $\Delta \nu$ and $\Delta P$, and then used the results to select out over two hundred thousand RCs, claiming a contamination of $\sim 9$ per cent.

%The current work
This paper is an update of H15.
The update is twofold.
First, in H15, RCs are selected from the second data release of LAMOST Galactic spectroscopic surveys.
Here the analysis is extended to the fourth release and thus yields more RCs that cover a larger Galactic disk volume.
Secondly, we attempt to derive the masses and ages of those selected RCs.
For low-mass giant stars, robust age can be estimated with mass well measured.
At present, age estimates of a precision of about 20 per cent (e.g. Martig et al. 2016a; Ness et al. 2016; Wu et al. 2018) have been achieved for red giants in the {\it Kepler} fields, along with very precise asteroseismic masses (better than 8-10 per cent; e.g. Huber et al. 2014; Yu et al. 2018) and atmospheric parameters from spectroscopy (e.g. the APOGEE and LAMOST surveys).
Those masses and ages, inferred from precise asteroseismic measurements, are then taken as training sets to derive masses and ages of over hundred thousand giant stars selected from the LAMOST and APOGEE surveys, using either an empirical relation based on the spectroscopic  carbon to nitrogen abundance ratios (e.g. Martig et al. 2016a; Ness et al. 2016; Sanders \& Das 2018) or a data-driven approach (e.g. Ness et al. 2016; Ho et al. 2017; Ting et al. 2018; Wu et al. 2019). 
The physics underlying the method of deriving masses and ages from the medium/high resolution spectra is that the carbon to nitrogen abundance ratio [C/N]  is tightly correlated with stellar mass, as a results of the convective mixing through the CNO cycle (aka the first-dredge up process).
In this sense, [C/N] ratios deducible from the optical/infrared spectra, are good indicators of stellar masses for red giants, and thus can be further used to derive their ages.
In the current work, we applied a similar technique used by {Wu et al. (2018, 2019)} to the selected RCs and derived their masses and ages, using a training set selected from the LAMOST-{\it Kepler} common stars (see Section\,4).
Moreover, atmospheric parameters, line-of-sight velocities and elemental abundances estimated from the LAMOST spectra (Xiang et al. 2017a,b), and proper motions from Gaia DR2 (Gaia Collaboration et al. 2018; Lindegren et al. 2018) are also provided for the RC sample.

The paper is structured as follows.
In Section\,2, we describe the data employed in the current analysis.
We then introduce the selection of primary RCs in Section\,3 and determine their masses and ages in Sections\,4.
In Section\,5, we present a new calibration of the $K_{\rm s}$ absolute magnitudes of primary RCs and derive their distances based on the new calibration.
The selection function of the primary RC sample is presented in Section\,6.
We present the final catalog of selected primary RCs in Section\,7.
Finally, a summary is given in Section\,8.

\section{Data}
\subsection{LAMOST Galactic surveys}
Being a 4 meter quasi-meridian reflecting Schmidt telescope equipped with 4000 fibers distributed in a field of view of about 20 sq. deg., LAMOST can simultaneously collect per exposure up to 4000 spectra covering the wavelength range 3800-9000\,\AA\, at a resolution of about 1800 (Cui et al. 2012). 
For the scientific motivations and target selections of the surveys, please refer to Zhao et al. (2012), Deng et al. (2012) and Liu et al. (2014) for details. 
After one-year Pilot Surveys between 2011 September and 2012 June, the five-year Phase-I LAMOST Regular Surveys began in 2012 September and completed in the summer of 2017.
Adding a new component of medium resolution ($R \sim 7500$) surveys, the Phase-II LAMOST Regular Surveys were initiated in 2019 September, following one-year Pilot Surveys between 2017 September and 2018 June.

To derive stellar atmospheric parameters and line-of-sight velocities from the LAMOST spectra, two independent stellar parameter pipelines have been developed -- the official LAMOST Stellar Parameter Pipeline (LASP; Luo et al. 2015) and the LAMOST Stellar Parameter at Peking University (LSP3; Xiang et al. 2015, 2017a). 
Typical uncertainties of 5\,km\,s$^{-1}$ in $v_{\rm los}$, 100\,K in  $T_{\rm eff}$, 0.25\,dex in log\,$g$ and 0.10\,dex in metallicity [Fe/H] have been achieved by both pipelines for `normal' type (FGK) stars.
Based on a {kernel principal component analysis}, the latest version of LSP3 (Xiang et al. 2017a) is able to deliver log\,$g$ of red giant stars with precision of about 0.1\,dex for red giant stars, using a training set from the LAMOST-{\it Kepler} common stars.
With the same technique, estimates of [$\alpha$/Fe], [C/H] and [N/H] of giant stars with precisions of 0.05,\,0.10 and 0.10\,dex , respectively, are also achieved, using a training set from the  LAMOST-APOGEE common stars.
Based on the derived atmospheric parameters, values of the interstellar extinction are also derived for the stars, using the `star pair' method developed by Yuan et al. (2013).
The precision of the estimated $E (B-V)$ values is about 0.04\,mag.
%Derived with the latest version of LSP3, the aforementioned parameters for stars surveyed between 2011 September and 2016 June are available at \url{http://dr4.lamost.org/doc/vac} (Xiang et al. 2017b).
In the current work, we adopt the atmospheric parameters (i.e., $T_{\rm eff}$, log\,$g$ and [Fe/H]) derived with LSP3 from the low resolution spectra ($R \sim 1800$) in the fourth data release of the LAMOST Galactic surveys (\url{http://dr4.lamost.org/doc/vac}; Xiang et al. 2017b).

\subsection{Asteroseismic samples from the Kepler data}
In addition to the LAMOST spectra and related catalogs of stellar parameters, {different asteroseismic samples constructed from the {\it Kepler} data are also used for training, examining and mass estimation purposes.}
Similar to H15, we use the asteroseismic sample with the evolutionary stages of the stars classified (based on $\Delta \nu$ and $\Delta P$), from Stello et al. (2013), to select primary RCs but excluding RGBs and secondary RCs.
%Most recently, Vrard et al. (2016) presented a much larger sample of  6100 red giant stars, 2 times larger than those of Stello et al. (2013), with evolutionary stages classified using the  {\it Kepler} data.
%We thus employ the newly presented stars from Vrard et al. (2016) for examination.
{We also have tested the selection of primary RCs trained by more recent asteroseismic samples (e.g., the sample of Vrard et al. 2016) but no significant improvements on the selection are found.}
Finally, we adopt the global asteroseismic parameters $\nu_{\rm max}$ and $\Delta \nu$ measured by Yu et al. (2018) for mass estimation.
Using a modified version of the SYD pipeline (Huber et al. 2009), {Yu et al. (2018)} have systematically characterized solar-like oscillations and granulation for 16\,094 oscillating red giants, based on the end-of-mission long-cadence data of {\it Kepler}.
The derived parameters are quite precise, with typical relative uncertainties of a few per cent.

\section{Selection of primary RCs}
As mentioned, to select a clean sample of primary RCs, we adopt the method developed in H15.
The method relies on the accurate estimation of surface gravity log\,$g$. 
Thanks to the training sample from LAMOST-{\it Kepler} fields (De Cat et al. 2015; Ren et al. 2016; Zong et al. 2018), a high precision (of about 0.1\,dex) has been achieved with LSP3 for the estimation of log\,$g$ from the LAMOST spectra {(Xiang et al. 2017a)}.
Such a precision of log\,$g$, together with a typical uncertainty of 100\,K of $T_{\rm eff}$, allow one to select primary RCs with a very low contamination (Bovy et al. 2014; H15; Ting et al. 2018).

The method of H15 includes two steps.
The first step is to select RCs in the metallicity-dependent $T_{\rm eff}$-log\,$g$ plane, supervised by an asteroseismic sample with the evolutionary stage classified by {Stello et al. (2013)}.
In principle, one can directly adopt the relation found by H15.
However, the values of $T_{\rm eff}$ and log\,$g$ used by H15 have been updated by the latest version of LSP3 {(Xiang et al. 2017a)}.
We therefore re-derive the relation using the updated $T_{\rm eff}$ and log\,$g$.
To do so, we cross match red giant stars from the LAMOST-{\it Kepler} fields with those of {Stello et al. (2013)}, and a total of 1547 common stars {(hereafter LE sample for short)} with LAMOST spectral SNR greater than 50 are found.
With the LE sample, we derive the following cuts in the metallicity-dependent $T_{\rm eff}$-log\,$g$ plane to select RC-like stars,
\begin{equation}
1.8 \leq {\rm log} g\,({\rm dex})\leq 0.0006\,{\rm K}^{-1} [T_{\rm eff} - T_{\rm eff}^{\rm Ref}] + 2.5\text{,}
\end{equation}
where 
\begin{equation}
T_{\rm eff}^{\rm Ref} = -873.1\,{\rm K\,dex}^{-1}{\rm [Fe/H]} + 4255\,{\rm K}\text{.}
\end{equation}
Similar in H15, the cuts are empirically obtained by maximizing the product of completeness and purity of the RC stars in the LE sample.
As an example, the distribution of LE sample stars of Solar metallicity ($-0.05 <$\,[Fe/H]\,$< 0.05$) in the $T_{\rm eff}$-log\,$g$ plane is presented in Fig.\,1.
Clearly, the cuts defined by above separate RC-like and RGB stars very well.
As in H15, the distribution of stars predicted by the PARSEC stellar isochrones (Bressan et al. 2012) is also shown as background in Fig.\,1.  
The prediction was generated assuming a constant star formation history for the last 10\,Gyr, a lognormal initial mass distribution from Chabrier (2001) and  a metallicity distribution similar to that of the LE sample stars for the individual metallicity bins.
The distribution of the LE sample stars is generally in agreement with the predicted one, except for possibly some minor systematic offsets between the LSP3 and PARSEC values of $T_{\rm eff}$. 

\begin{figure}
\begin{center}
\includegraphics[scale=0.35,angle=0]{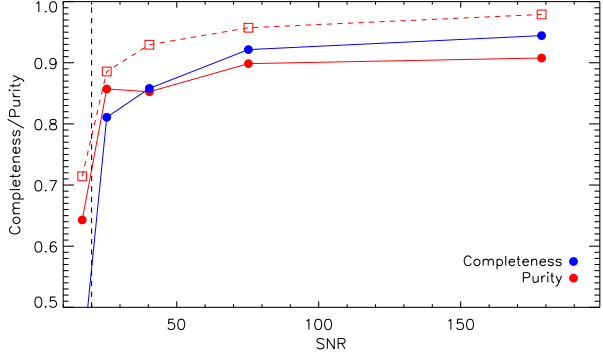}
\caption{Completeness (blue dots) and purity (red dots) of the primary RC sample constructed with the method described in Section\,3, as deduced using the asteroseismic sample with evolutionary stages classified by {Vrard et al. (2016)}. The purity as a function of SNR of all RCs (primary plus secondary RCs) is also shown by red squares.
{The black dashed line marks spectral SNR\,$= 20$.}}
\end{center}
\end{figure}

In H15, a second step was developed to {remove secondary RCs}, using cuts in the color-metallicity plane. 
Since the metallicities [Fe/H] yielded by the latest version of LSP3 change little compared to earlier results, we have directly adopted the cuts obtained by H15,
\begin{equation}
Z < 2.58[(J - K_{\rm s})_0 - 0.400]^3 + 0.0034\text{,}
\end{equation}
and,
\begin{equation}
Z > 1.21[(J - K_{\rm s})_0 - 0.085]^9 + 0.0011\text{.}
\end{equation} 
Here the values of intrinsic color $(J-K_{\rm s})_0$ are from the 2MASS survey (Skrutskie et al. 2006), after correcting for reddening estimated with the `star pair' method (see Section\,2.1; Yuan et al. 2013).
Values of metallicity $Z$ are converted from those of [Fe/H] using the relation given by Bertelli et al. (1994) assuming a Solar value $Z_{\odot} = 0.017$.
As shown by Fig.\,1, the cuts defined by Eqs.\,(3) and (4) can efficiently eliminate secondary RCs.
The latter are descendants of high-mass ($\ge 2 M_{\odot}$) in helium burning phase, ignited non-degenerately.
They are non-standard candles. 
In addition to the above cuts developed in H15, we consider only stars of $T_{\rm eff} \leq 5000$\,K and log\,$g \le 2.75$ to further suppress the contamination of secondary RCs and RGBs {(see the black pluses in the right panel of Fig.\,1)}. 

By applying above cuts to the whole set of LAMOST Data Release 4 processed with the latest version of LSP3 (LMDR4 hereafter; Xiang et al. 2017b), about 150,000 primary RC candidates of spectral SNR\,$> 10$ are obtained.
To characterise the purity and completeness of the selected primary RC sample, we use another asteroseismic sample with the evolutionary stage classified by {Vrard et al. (2016)}. 
The sample contains more than 6100 stars.
To do so, we first exclude stars in the sample of Vrard et al. (2016) that are in common with the LE sample and cross match the remaining ones with the $\sim 150,000$ selected primary RC candidates and also with the whole LMDR4.  
This yields 1720 and 2836 common stars, respectively.
We divide the 1720 common stars into different SNR bins and calculate the numbers of primary RCs ($N_{\rm SpRC}$), secondary RCs ($N_{\rm SsRC}$) and RGBs ($N_{\rm SRGB}$) in each SNR bin.
Similarly, the 2836 common stars are also divided into the same SNR bins and the numbers of primary RCs ($N_{\rm ApRC}$), secondary RCs ($N_{\rm AsRC}$) and RGBs ($N_{\rm ARGB}$) in the individual bins are derived.
The purity and completeness for the individual spectral SNR bins are then simply given by ratios $N_{\rm SpRC}$/($N_{\rm SpRC}$ + $N_{\rm SsRC}$ + $N_{\rm SRGB}$) and $N_{\rm SpRC}$/$N_{\rm ApRC}$, respectively.
The results are presented in Fig.\,2.
The purity is around 65 per cent for a SNR of 10 and increases rapidly to about $85$ per cent for SNRs higher than 20.
The completeness is low, of about 40 per cent at SNR\,$\sim 10$ and increases rapidly to over 85 per cent for SNR\,$> 50$.
Based on the results, we further exclude primary RC candidates of SNR\,$< 20$. 
This leaves with us nearly 140,000 stars.
Over 87 per cent of the remaining stars have SNRs higher than 30, implying a purity and a completeness exceeding $80$ per cent for most of our primary RCs in the final sample.
If only concerning RCs that include both primary and secondary RCs, our technique is expected to deliver a sample of purity over 95 per cent at SNR\,$> 50$ (see the squares in Fig.\,2), comparable to the recent results of Ting et al. (2018).
{Here, the above examination performance could be decreased, if any systematic differences between the training sample of Stello et al. (2013) and the test sample of Vrard et al. (2016).
In this degree, the reported purity and completeness in Fig.\,2 are actually lower bounds.}

\begin{figure}
\begin{center}
\includegraphics[scale=0.475,angle=0]{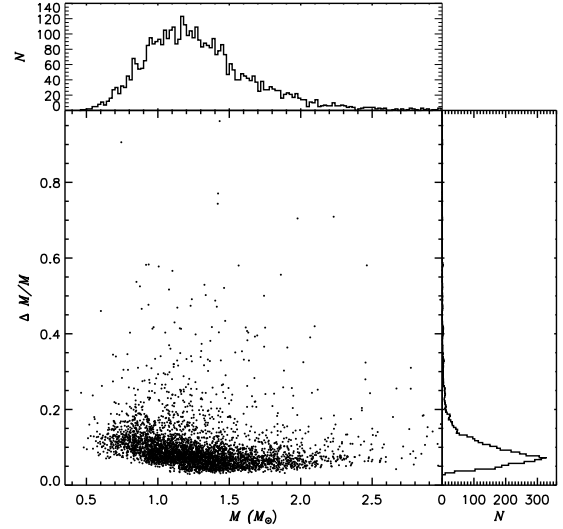}
\caption{Distributions of masses of primary RCs estimated from the asteroseismic and spectroscopic stellar atmospheric parameters {(see Section\,4.1 for more details)}, and of the associated errors.}
\end{center}
\end{figure}

\begin{figure*}
\begin{center}
\includegraphics[scale=0.45,angle=0]{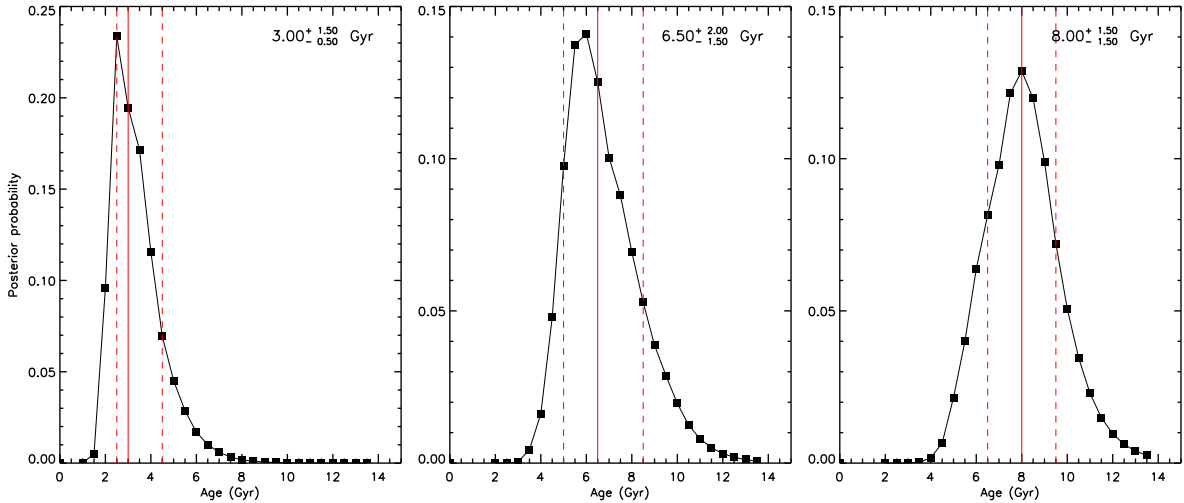}
\caption{Posterior probability distributions of stellar ages given by our Bayesian approach for three example stars of different ages. 
              {The details of the approach are presented in Section\,4.2, Xiang et al. (2017c).}
              The red solid and dashed lines denote the medians and the 68 per cent probability intervals of the stellar ages, respectively.}
\end{center}
\end{figure*}

\begin{figure}
\begin{center}
\includegraphics[scale=0.475,angle=0]{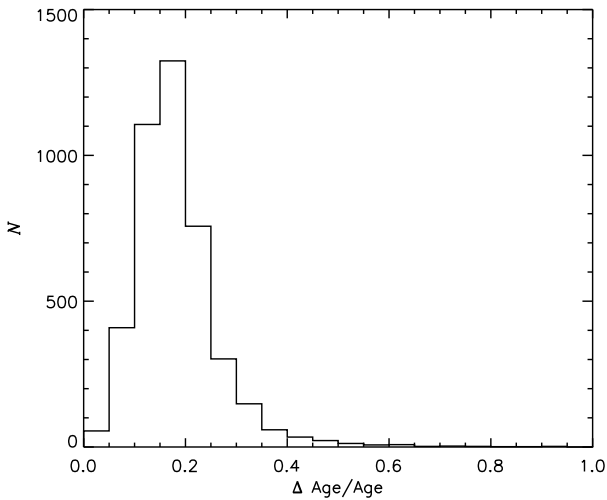}
\caption{Distribution of the relative uncertainties of ages estimated {in Section\,4.1 for the training stars.}}
\end{center}
\end{figure}

\section{Masses and Ages of primary RCs}

\begin{figure*}
\begin{center}
\includegraphics[scale=0.55,angle=0]{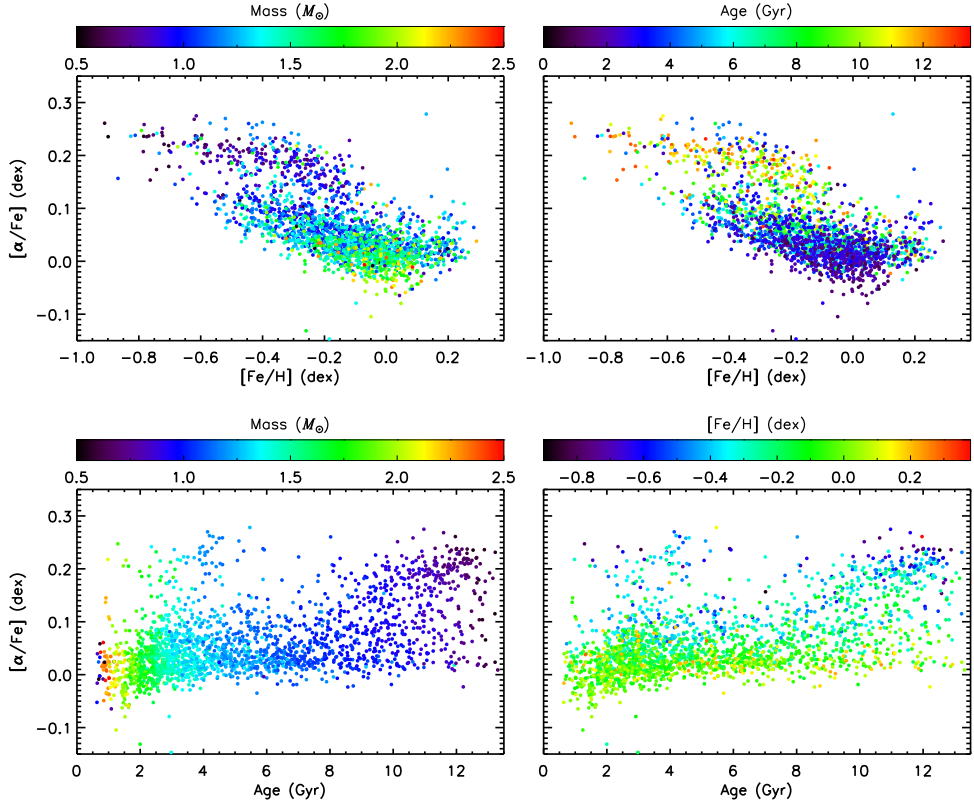}
\caption{{\it Upper panels:} Distributions of  stars in the training sample in the [Fe/H]--[$\alpha$/Fe] plane, color-coded by stellar mass (left panel) and age (right panel).
{\it Lower panels:} Distributions of stars in the training sample in the age--[$\alpha$/Fe] plane, color-coded by stellar mass (left panel) and metallicity (right panel).
{Here the values of [Fe/H] and [$\alpha$/Fe] are those estimated with LSP3.
The training sample is defined in Section\,4.1.}}
\end{center}
\end{figure*}

\begin{figure*}
\begin{center}
\includegraphics[scale=0.4,angle=0]{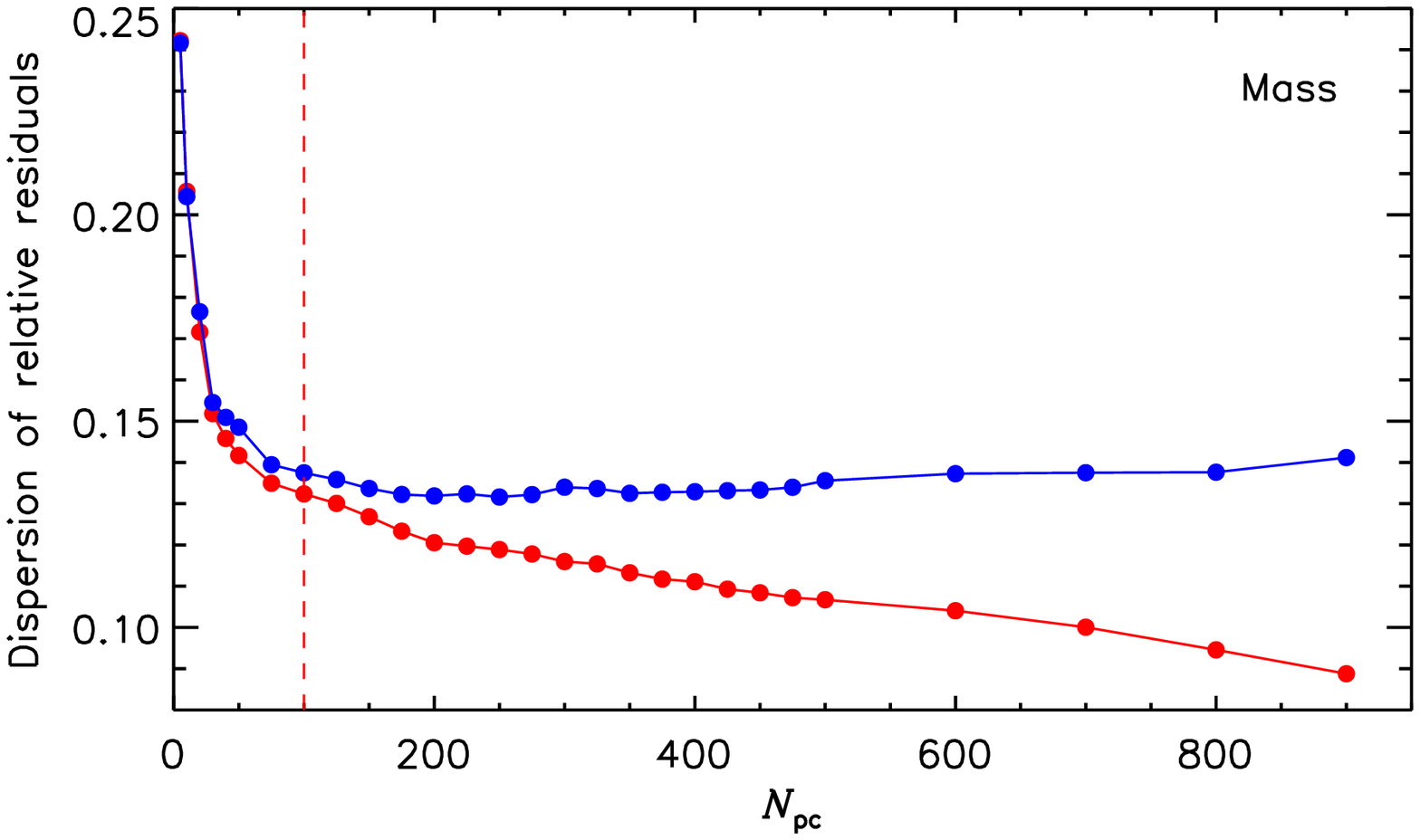}
\includegraphics[scale=0.4,angle=0]{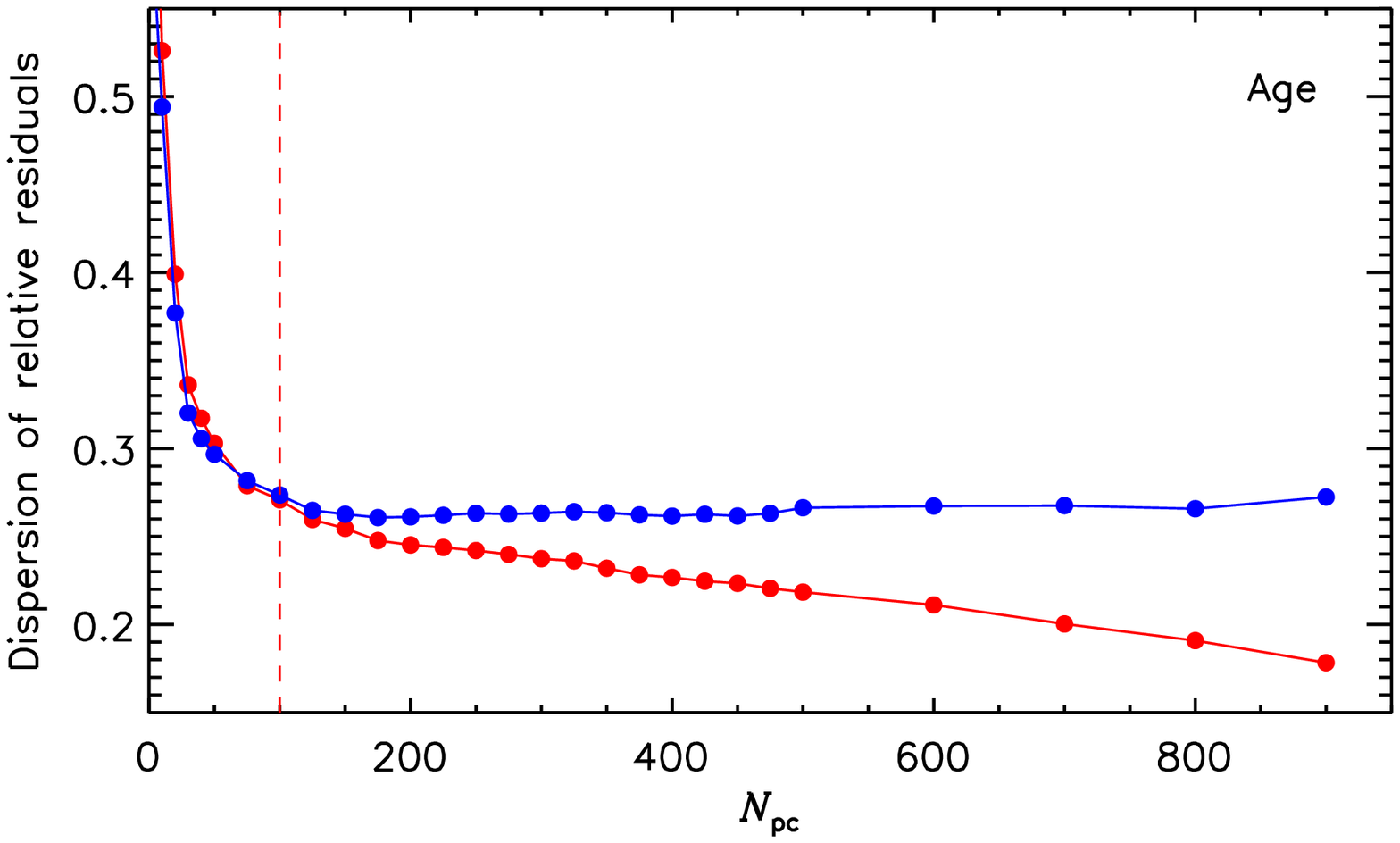}
\caption{The dispersions of the relative residuals of deduced masses (left) and ages (right) for the training (red dots) and {test} (blue dots) sets as a function of $N_{\rm PC}$. The plots suggest that the optimal value of $N_{\rm pc}$ for training stellar mass and age (see Section\,4.2) is about 100 (see the red dashed lines in both panels).}
\end{center}
\end{figure*}

\begin{figure*}
\begin{center}
\includegraphics[scale=0.4,angle=0]{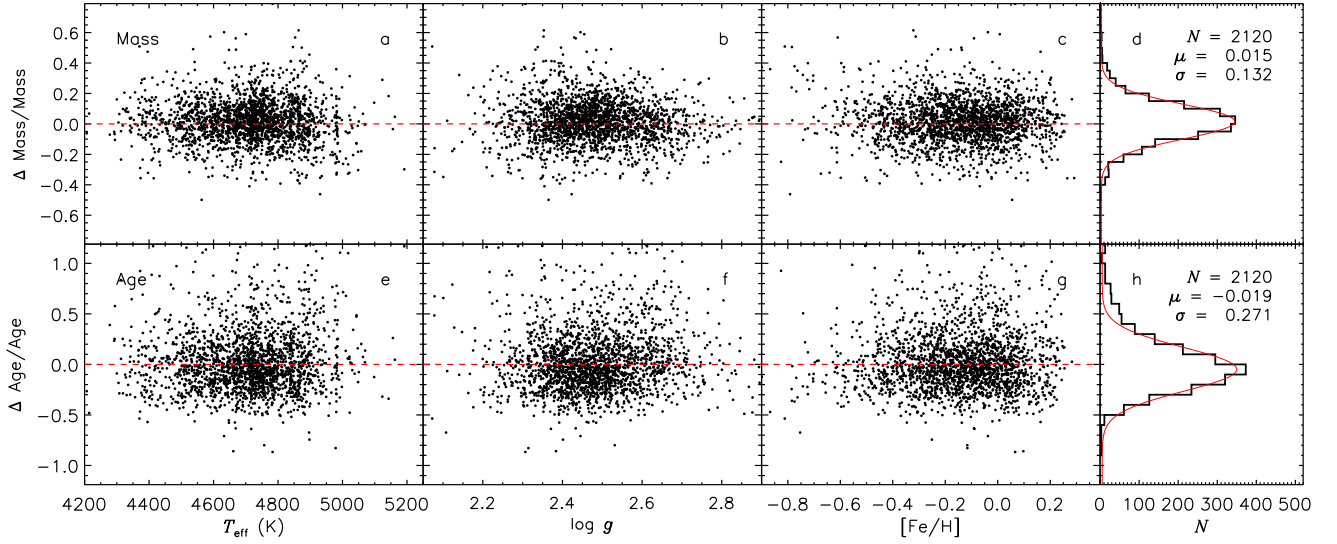}
\caption{Distributions of relative mass (upper panel) and age (lower panel) residuals for the training sample as a function of LSP3 $T_{\rm eff}$ (panels a and e), log\,$g$ (panels b and f) and [Fe/H] (panels c and g).
Panels d and h show histograms of the residuals (black line). Also overplotted in red is a Gaussian fit to the histogram distribution.
For the current KPCA analysis, $N_{\rm PC} = 100$ is adopted.}
\end{center}
\end{figure*}

\begin{figure*}
\begin{center}
\includegraphics[scale=0.475,angle=0]{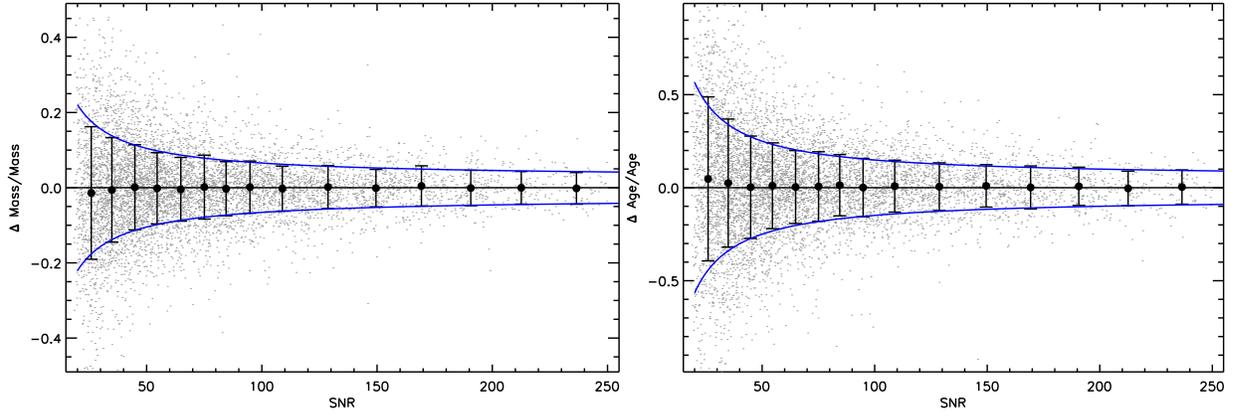}
\caption{Relative internal residuals of mass (left) and age (right) given by duplicate observations of different spectral SNRs.
Here only duplicate observations of SNR differences smaller than 20 per cent are adopted and their means are assumed as the final SNRs.
Black dots and error bars represent the medians and standard deviations (after divided by $\sqrt{2}$) of the relative residuals in the individual  spectral SNR bins.
Blue lines indicate fits of the standard deviations as a function of spectral SNR, as given in Eq.\,(8).}
\end{center}
\end{figure*}

\begin{figure*}
\begin{center}
\includegraphics[scale=0.6,angle=0]{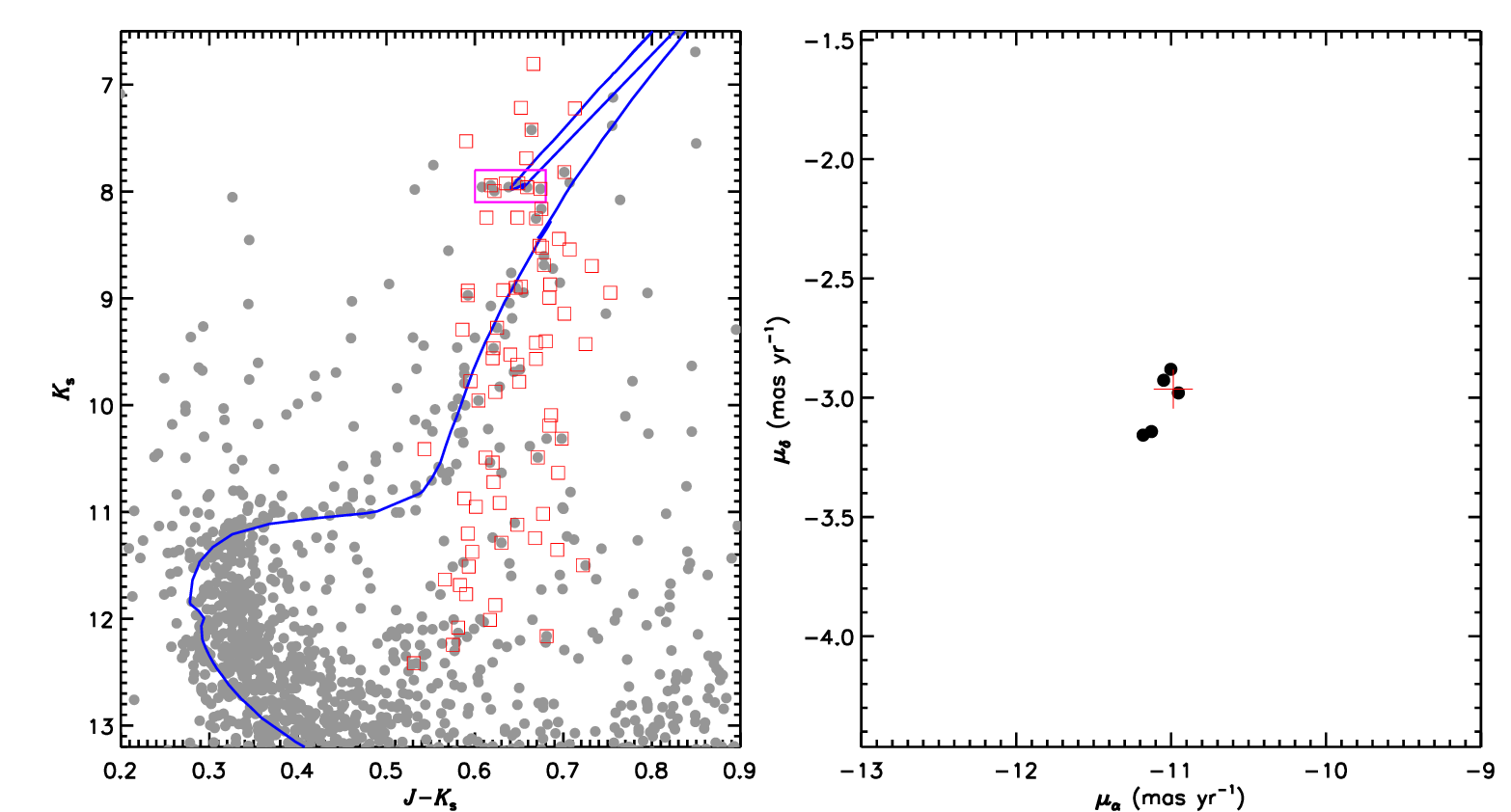}
\caption{Selection of member candidates of OC NGC\,2682, based on positions in the CMD (left) and proper motions (right).
{\it Left panel:} The grey dots represent stars within 1.5 cluster diameter from the cluster center.
The red squares denote primary RCs within four cluster diameters from the cluster center that have 2MASS photometry.
The blue line is an isochrone from PARSEC with age of 4\,Gyr and Solar metallicity, after adding a distance modulus and corrected for the interstellar reddening with information in the literature (see Table\,2). 
The magenta box indicates the location of stars in the RC evolutionary stage for this cluster. 
{\it Right panel:} Proper motion diagram showing the spectroscopic candidates within in the magenta box shown in the left panel.
The red plus represents the cluster proper motions from Gaia DR2 (Cantat-Gaudin et al. 2018; see Table\,2).}
\end{center}
\end{figure*}

\subsection{Training and test sets}
In this Section, we estimate the masses and ages for our selected primary RCs.
{To do so}, we first construct a primary RC sample with precise mass and age values deduced using the asteroseismic information provided by {Yu et al. (2018)} and the stellar atmospheric parameters from LAMOST.
We then cross-match the sample {of Yu et al. (2018)} with our primary RC sample. 
A total of 4285 common stars are found.
Masses of those common stars are determined using the standard seismic scaling relation,
\begin{equation}
\frac{M}{M_{\odot}} = (\frac{\Delta \nu}{\Delta \nu_{\odot}})^{-4}(\frac{\nu_{\rm max}}{\nu_{{\rm max}, \odot}})^3(\frac{T_{\rm eff}}{T_{{\rm eff}. \odot}})^{3/2}\text{,}
\end{equation}
Here we adopt Solar values $T_{{\rm eff}, \odot} = 5777$\,K, $\nu_{{\rm max}, \odot} = 3090\,\mu$Hz and $\Delta \nu_{\odot} = 135.1\,\mu$Hz (Huber et al. 2011).
However, as pointed out by previous studies (e.g. Huber et al. 2011; Viani et al. 2017), the standard scaling relation may induce systematic errors of 10 to 15 per cent in the derived masses.
Similar to {Wu et al. (2019)}, we adopt a modified scaling relation from Sharma et al. (2016) to reduce the systematic errors.
Based on the theoretical stellar models, a correction factor $f_{\Delta \nu}$ is added in the standard scaling relation,
\begin{equation}
\frac{M}{M_{\odot}} = (\frac{\Delta \nu}{f_{\Delta \nu}\Delta \nu_{\odot}})^{-4}(\frac{\nu_{\rm max}}{\nu_{{\rm max}, \odot}})^3(\frac{T_{\rm eff}}{T_{{\rm eff}, \odot}})^{3/2}\text{,}
\end{equation}
where $f_{\Delta \nu}$ can be obtained for all stars using the publicly available code {\it Asfgrid}\footnote{\url{http://www.physics.usyd.edu.au/k2gap/Asfgrid/}} provided by {Sharma et al. (2016)}.
With this modified scaling relation, masses of the 4285 common stars are derived using $\Delta \nu$ and $\nu_{\rm max}$ provided in {Yu et al. (2018)} and $T_{\rm eff}$ yielded by the LSP3.
As Fig.\,3 shows, masses thus derived for those primary RCs largely fall between 0.4 and 2.0 $M_{\odot}$, except for a few of masses greater than 2\,$M_{\odot}$, probably from contamination of secondary RCs.

With the masses estimated above and the spectroscopic stellar atmospheric parameters ($T_{\rm eff}$, log\,$g$ and [Fe/H]), the ages of those primary RCs can be further derived using the stellar isochrones.
{To do so}, we have adopted a Bayesian approach similar to Xiang et al. (2017c).
The input constraints include the mass $M$ and surface gravity log\,$g$ inferred from asteroseismic parameters, and the effective temperature $T_{\rm eff}$ and metallicity [Fe/H] estimated from the LAMOST spectra with LSP3.
For the stellar isochrones, we have adopted the PARSEC ones calculated with a mass-loss parameter $\eta_{\rm Reimers} =0.2$ (Bressan et al. 2012).
Details of age estimation with Bayesian approach are described by Xiang et al. (2017c).
To show the {results} of our age determinations, posterior probability distributions as a function of age for three stars of typical ages are shown in Fig.\,4.
Thanks to the very precise mass estimates from the asteroseismology information, the distributions show prominent peaks, yielding well constrained ages.
As Fig.\,5 shows, the typical uncertainties of the estimated ages are about 15-20 per cent, except for a few (about 3 per cent) of uncertainties larger than 40 per cent (as a result of the large errors in the derived mass and/or in the atmospheric parameters). 

With masses and ages estimated for those 4285 stars, we then divide them into two sub-samples, a training and a test one.
For the training sample, 2120 stars are randomly picked out with a spectral SNR greater than 50, a mass error smaller than 15 per cent and an age error smaller than 40 per cent.
In Fig.\,6, the distributions of the training sample  in the [Fe/H]--[$\alpha$/Fe] plane, color coded by mass and age, and  in the age--[$\alpha$/Fe] plane, color coded by mass and metallicity, are shown.
We note that the discarded stars do not change the distributions of mass, age, [Fe/H] and [$\alpha$/Fe] of the test set.
The remaining 987 stars of mass errors smaller than 15 per cent and age error smaller than 40 per cent are adopted as the test set.

\subsection{Mass and age determinations for primary RCs}
In this Section, we estimate masses and ages from the LAMOST spectra, using the Kernel Principal Analysis (KPCA; Sch{\"o}lpokf et al. 1998) method trained with the training sample described above.
Generally, PCA is a classic and powerful method that can convert observations (e.g. spectra) into a set of linearly uncorrelated orthogonal variables or principal components (PCs).
KPCA works like PCA but can be extended to nonlinear feature extractions with kernel techniques (Gaussian radial basis function is adopted here).
The approach has been successfully used to estimate stellar atmospheric parameters, as well as masses and ages  from the LAMOST spectra by H15, {Xiang et al. (2017a) and Wu et al. (2019)}, respectively.
For details of this method, please refer to those papers.

Similar to {Xiang et al. (2017a)},  the LAMOST blue-arm spectra ($3900$--5500\,\AA) are adopted to train the relations between the spectral features and stellar mass and age.
The key of the training is to find an optimal number of principal components ($N_{\rm PC}$).
Generally, a small value of $N_{\rm PC}$ is not sufficient to construct tight relations between the spectral features and the parameters to be estimated (stellar masses and ages here). 
On the other hand, an excessively large value  of $N_{\rm PC}$ has the risk of over-fitting and causes problems when dealing with spectra of relatively low SNRs.

To determine an optimal value of $N_{\rm PC}$, we have tried different values of $N_{\rm PC}$ and compared the dispersions of the relative residuals of the derived masses and ages for both training and {test} sets.
For the training sample, the dispersions of the relative residuals of the deduced masses and ages decrease monotonically from 0.24 to 0.09 and from 0.56 to 0.18, respectively, as $N_{\rm PC}$ increases from 5 to 900 (see Fig.\,7).
However, for the test set, the dispersion of the relative mass residuals decreases from 24 to 13 per cent as $N_{\rm PC}$ increases from 5 to 125, and then remains unchanged as $N_{\rm PC}$ continuously increases to 900. 
Similar result is also found for the relative age residuals (see Fig.\,7).
Obviously, for $N_{\rm PC} \ge 125$, the algorithm over-fits the training set for both stellar masses and ages.
Finally, the turning point, $N_{\rm PC} = 100$, in the dispersions of relative mass and age differences as a function of $N_{\rm PC}$ (as shown in Fig.\,7) is adopted to train the relations between the spectral features and stellar masses and ages.
The relative residuals of masses and ages, given by the training set for $N_{\rm PC} = 100$, as a function of atmospheric parameters ($T_{\rm eff}$, log\,$g$ and [Fe/H]) are shown in Fig.\,8 and no obvious trends of variations with those parameters  are detected.
The typical dispersions of the relative residuals are 13 and 27 per cent, respectively.

\begin{table}
\centering
\caption{Fit coefficients of the mass and age uncertainty estimates}
\begin{tabular}{ccc}
\hline
Coeff.&Mass&Age\\
\hline
$a$&0.0258&0.0541\\
$b$&0.9879&1.0475\\
$c$&3.7579&11.7923\\
\hline
\end{tabular}
\end{table}

\begin{table*}
\centering
\caption{Comparison of KPCA ages, as well as other LSP3 stellar parameters, with the literature values for open clusters}
\begin{threeparttable}
\begin{tabular}{cccccccc}
\hline

Parameter&NGC\,6811&NGC\,2420&NGC\,6819&NGC\,2682&NGC\,188&NGC\,6791&Be\,17\\
&&&&(M\,67)&&\\
\hline
\multicolumn{8}{c}{Literature} \\

$l$\tnote{a}&\,079.210&\,198.107&\,073.984&\,215.696&\,122.843&\,069.958&\,175.646\\

$b$\tnote{a}&$+12.015$&$+19.634$&$+08.491$&$+31.896$&$+22.384$&$+10.904$&$-03.648$\\

Diameter (arcmin)\tnote{a}&14&5&13&25&17&10&7\\

$E(B-V)$&$0.07\pm0.02$&$0.05$&$0.16\pm0.01$&$0.10\pm0.04$&$0.09\pm0.10$&$0.11\pm0.01$&$0.61\pm0.01$\\

Age (Gyr)&$1.00\pm0.10$&$2.20\pm0.30$&$2.38\pm0.22$&$4.00\pm0.60$&6.20$\pm$0.20&$9.50\pm0.30$&$10.06\pm2.77$\\

 [Fe/H]&$0.04\pm0.01$&$-0.20\pm0.06$&$-0.02\pm0.02$&$0.00\pm0.06$&$-0.03\pm0.04$&$0.42\pm0.05$&$-0.10\pm0.09$\\
  
RV (km\,s$^{-1}$)&$6.68\pm0.08$&$73.60\pm0.60$&$2.34\pm0.05$&$33.64\pm0.03$&$-42.36\pm0.04$&$-47.40\pm0.13$&$-73.70\pm0.80$\\

$\mu_{\alpha}$ (mas\,yr$^{-1}$)\tnote{b}&$-3.40 \pm 0.12$&$-1.19 \pm 0.14$&$-2.92 \pm 0.13$&$-10.99 \pm 0.19$&$-2.31 \pm 0.14$&$-0.42 \pm 0.17$&$2.62 \pm 0.28$\\

$\mu_{\delta}$ (mas\,yr$^{-1}$)\tnote{b}&$-8.81 \pm 0.12$&$-2.13 \pm 0.13$&$-3.86 \pm 0.14$&$-2.96 \pm 0.20$&$-0.96 \pm 0.15$&$-2.27 \pm 0.19$&$-0.35 \pm 0.18$\\

References\tnote{c} &1--2&3--4&5--8&9--12&4, 13--14&10,\,15--16&3, 17--18\\

\hline 
\multicolumn{8}{c}{This work} \\

$N_{\rm member}$&1&1&9&5&3&1&2\\

$E(B-V)$&$0.03\pm0.04$&$0.02\pm0.04$&$0.09\pm0.04$&$0.07\pm0.04$&$0.11\pm0.04$&$0.06\pm0.04$&$0.62\pm0.01$\\

KPCA Age (Gyr)&$0.99\pm0.28$&$2.75\pm0.80$&$2.68\pm0.25$ &$2.94\pm0.36$&$5.91\pm0.94$&$12.45\pm3.40$&$8.58\pm1.65$\\

[Fe/H]&$-0.05\pm0.11$&$-0.22 \pm 0.15$&$0.00\pm0.04$&$-0.04\pm0.04$&$-0.03\pm0.07$&$0.10\pm0.14$&$-0.18\pm0.07$\\

RV (km\,s$^{-1}$)&$0.38\pm3.32$&$67.40\pm4.23$&$2.81\pm1.16$&$35.46\pm1.39$&$-45.20\pm2.44$&$-47.40\pm0.13$&$-75.69\pm2.20$\\
\hline
\end{tabular}
\begin{tablenotes}
\item[a]From Dias  et  al.  (2012). 
\item[b]From Cantat-Gaudin et al. (2018).
\item[c] References: (1)\,Sandquist et al. (2016); (2)\,Molenda-{\.Z}akowicz et al. (2014); (3)\,Salaris, Weiss, \& Percival (2004); (4)\,Jacobson, Pilachowski, \& Friel (2011);
                                 (5)\,Anthony-Twarog et al. (2014); (6)\,Lee-Brown et al. (2015); (7)\,Brewer et al. (2016); (8)\,Hole et al. (2009); (9)\,An et al. (2007); (10)\,Heiter et al. (2014); 
                                 (11)\,Stello et al. (2016); (12)\,Geller, Latham, \& Mathieu (2015); (13)\,Meibom et al. (2009); (14)\,Geller et al. (2008); (15)\,An et al. (2015); 
                                 (16)\,Tofflemire et al. (2014); (17)\,Bragaglia et al. (2006); (18)\,Friel, Jacobson, \& Pilachowski (2005)
\end{tablenotes}
\end{threeparttable}
\end{table*}

\begin{figure}
\begin{center}
\includegraphics[scale=0.525,angle=0]{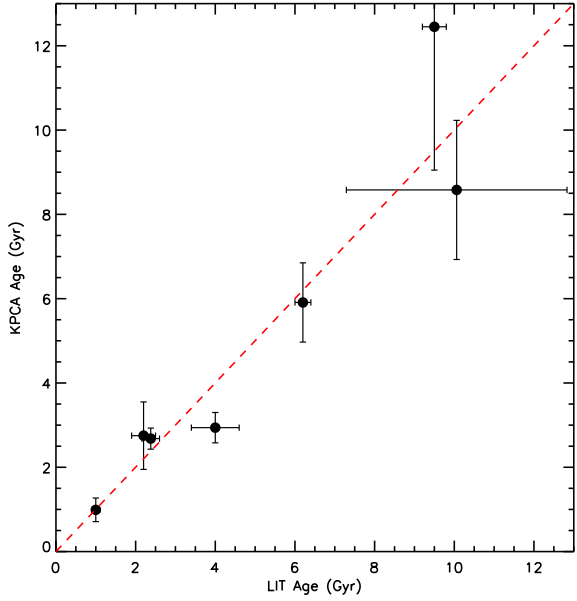}
\caption{Comparison of KPCA ages of OCs with literature values.}
\end{center}
\end{figure}

\begin{figure}
\begin{center}
\includegraphics[scale=0.325,angle=0]{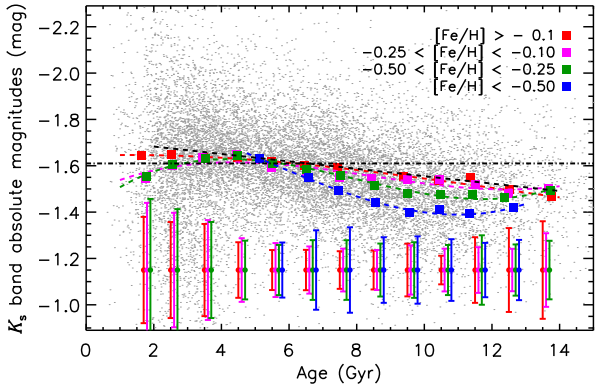}
\caption{$K_{\rm s}$ band absolute magnitudes of primary RCs as a function of age.
Squares of different colors represent the median values of $M_{K_{\rm s}}$ in the individual age bins for the various [Fe/H] ranges (red: [Fe/H]\,$> -0.10$, magenta: $-0.25 <$\,[Fe/H]\,$< - 0.10$, green: $-0.50 <$\,[Fe/H]\,$< - 0.25$ and blue square: [Fe/H]\,$< - 0.50$).
The error bars indicate the standard deviations (after subtraction of contributions from the photometric and astrometric uncertainties) {of $K_{\rm s}$ band absolute magnitudes} for the individual bins.
The dashed lines of different colors represent third-order polynomial fits to the squares of the corresponding colors.
The black dot-dashed  and dashed lines represent the calibrations of $K_{\rm s}$ absolute magnitudes of RCs by L12 and Chen et al. (2017), respectively.}
\end{center}
\end{figure}

\begin{figure}
\begin{center}
\includegraphics[scale=0.325,angle=0]{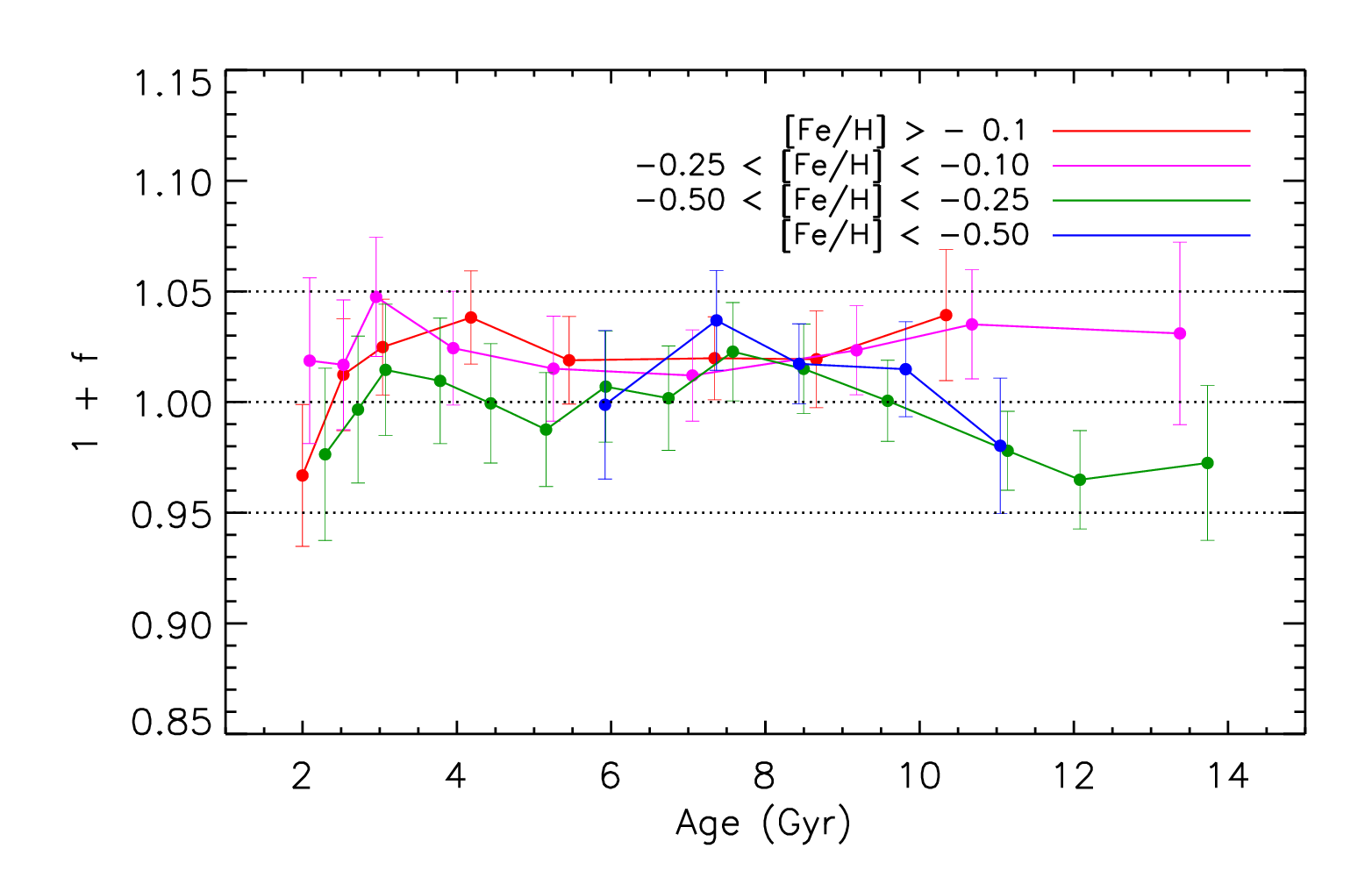}
\includegraphics[scale=0.325,angle=0]{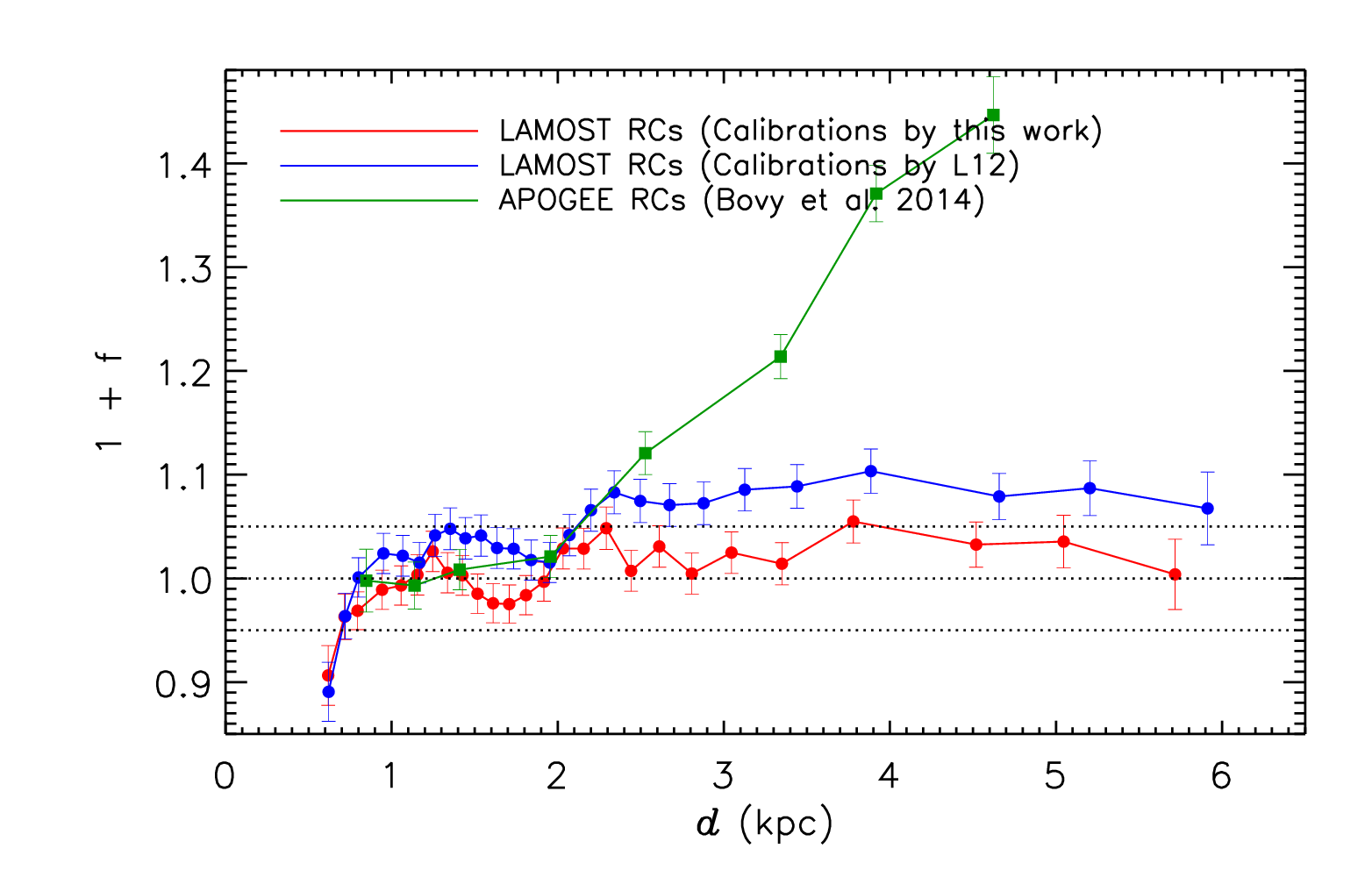}
\caption{Validation of the distances of primary RCs (see Section\,5.2).
{\it Upper panel:} relative error versus age for the LAMOST primary RCs of SNR\,$> 60$. 
The stars are grouped in age such that each bin contains 6000 stars, and the age mask slides in steps of 2000 stars. 
For the definition of estimator $1 + f$,  please refer to SBA12 for details. 
$f$ is zero for no biases in the estimated distances.
Here symbols of different colors indicate different metallicity ranges.
{\it Lower panel:} relative error versus distance for the different primary RC samples.
The red dots represent the LAMOST primary RC sample of SNR\,$> 60$ and distances estimated by the calibration reported in the current work.
The blue dots represent the LAMOST primary RC sample of SNR\,$> 60$ and distances estimated with the calibration of L12.
The green squares represent the APOGEE DR14 primary RC sample of SNR\,$> 60$ and distances estimated by Bovy et al. (2014).
The binning scheme is the same as for the upper panel.}
\end{center}
\end{figure}

\begin{figure}
\begin{center}
\includegraphics[scale=0.35,angle=0]{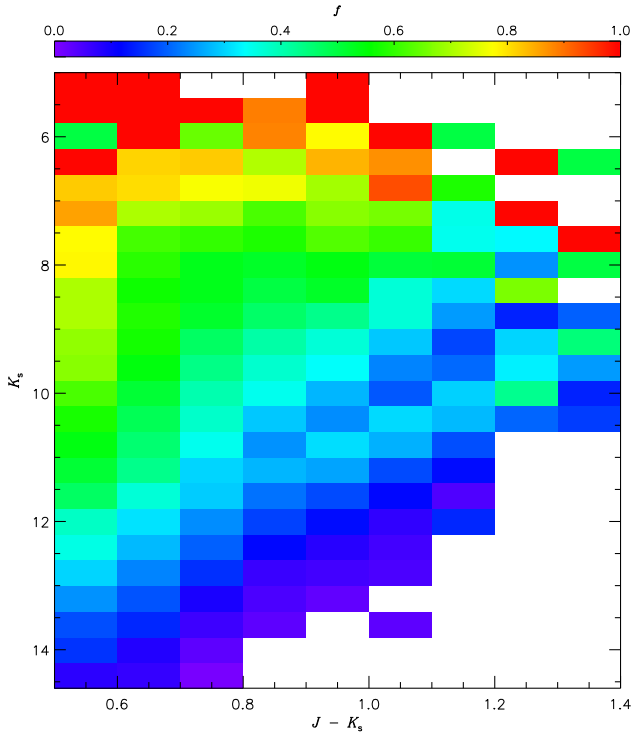}
\caption{Distribution of the mean selection fractions in the $J - K_{\rm s}$ versus $K_{\rm s}$ plane for our primary RC sample (see Section\,6).}
\end{center}
\end{figure}

\begin{figure*}
\begin{center}
\includegraphics[scale=0.425,angle=0]{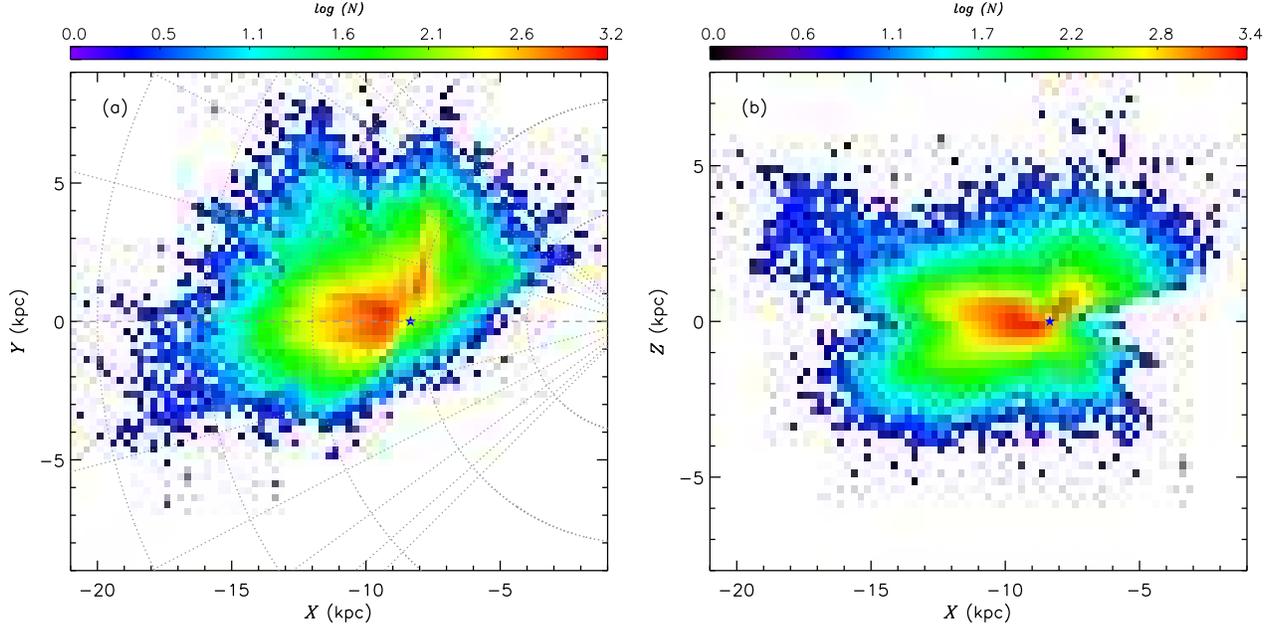}
\caption{Number density distribution of our primary RC sample stars in the $X$-$Y$ (left panel) and $X$-$Z$ (right panel) planes. The Sun (marked by the blue star in each panel) is at ($X$, $Y$, $Z$) = ($-8.34$, 0.00, 0.00)\,kpc.
The stars are binned by $0.25 \times 0.25$\,kpc$^2$. The densities are shown on a logarithmic scale.
The dashed grey lines (from top to bottom) in the left panel mark azimuthal angle $\phi$ from 75 to $-75^{\circ}$  in step of 15$^{\circ}$.
The grey dashed arcs (from right to left) in the left panel mark the projected Galactocentric radius $R$ from 4 to 20\,kpc in step of 4\,kpc.}
\end{center}
\end{figure*}

\begin{figure}
\begin{center}
\includegraphics[scale=0.375,angle=0]{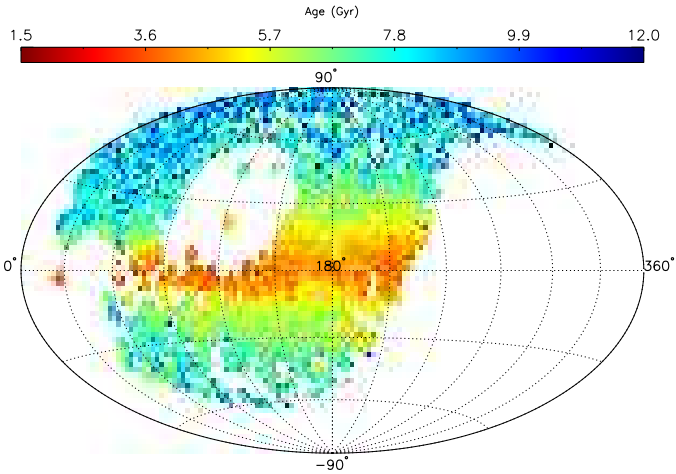}
\caption{Distribution of the mean stellar ages of our primary RC sample in the Galactic coordinate system. The data are divided into patches of $2.5^{\circ} \times 2.5^{\circ}$.}
\end{center}
\end{figure}

\begin{figure*}
\begin{center}
\includegraphics[scale=0.475,angle=0]{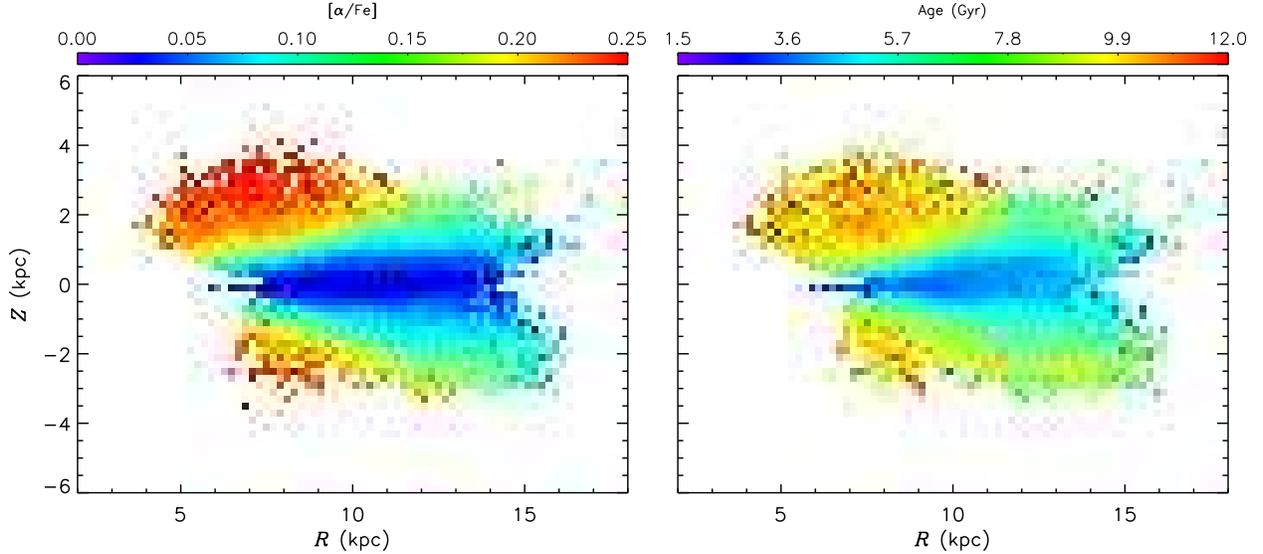}
\caption{Color-coded distributions of the mean values of [$\alpha$/Fe] (left panel) and stellar age (right panel) in the $R$-$Z$ plane.
The stars are binned by $0.20 \times 0.20$\,kpc$^2$. }
\end{center}
\end{figure*}

\begin{figure*}
\begin{center}
\includegraphics[scale=0.425,angle=0]{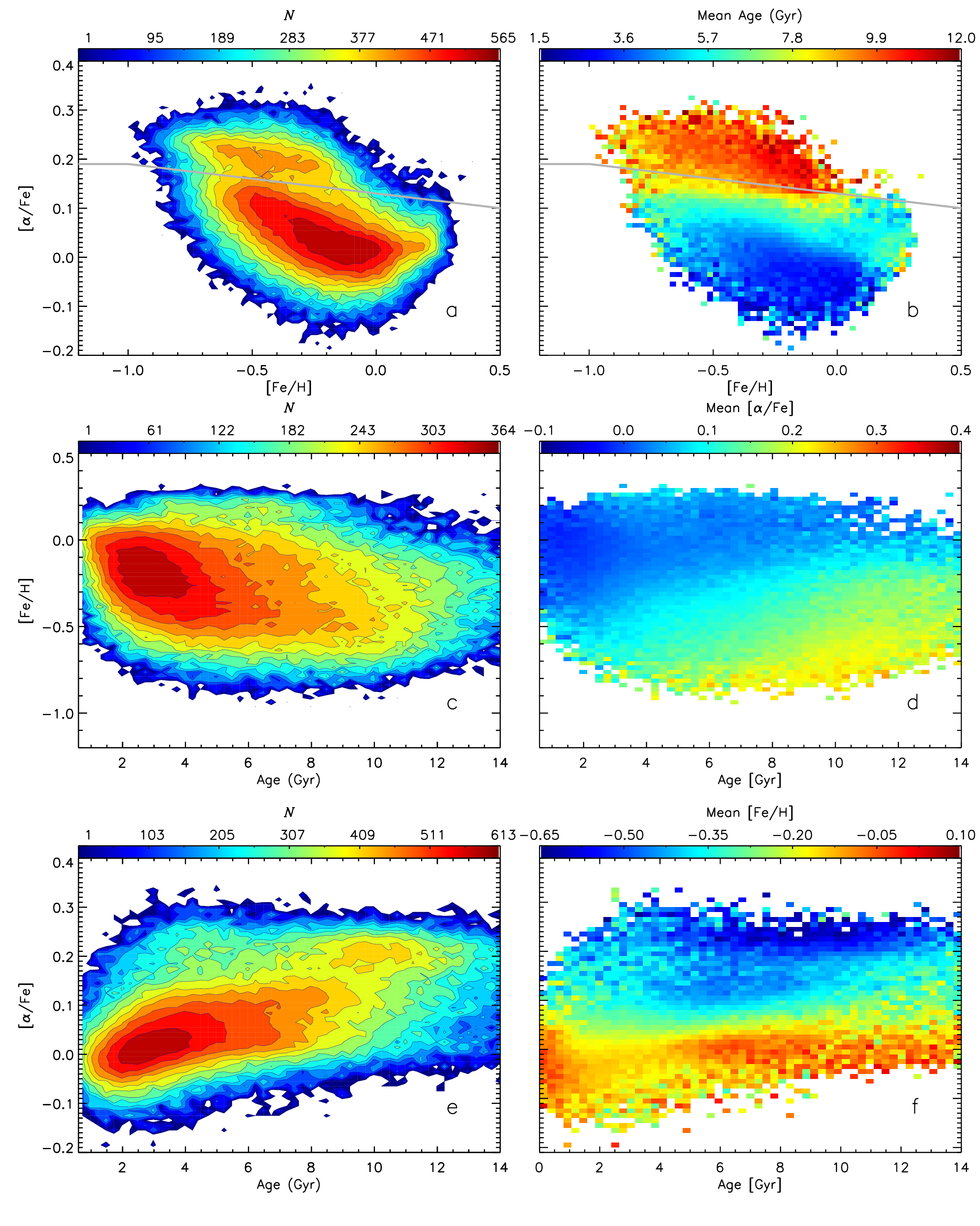}
\caption{Distributions of our RC sample stars in the i) [Fe/H]--[$\alpha$/Fe] plane, binned by 0.025 dex by 0.01 dex  and color-coded by number density (Panel a) and mean age (Panel b); ii) [Fe/H]--Age plane, binned by 0.25\,Gyr by 0.025\,dex and color-coded by number density (Panel c) and mean [$\alpha$/Fe] (Panel d); iii) [$\alpha$/Fe]--Age plane, binned by 0.25\,Gyr by 0.01\,dex and color-coded by number density (Panel e) and mean [Fe/H] (Panel f).
The grey line in Panels a and b represent an empirical cut to separate the chemical thin and thick disk stars. 
Stars with [$\alpha$/Fe]\,$\ge 0.125$ and age younger than 5\,Gyr (the so-called young $\alpha$-enhanced stars, see Section\,7.3 for details) are discarded when drawing Panel b.}
\end{center}
\end{figure*}

Using the relations as trained above, we have derived masses and ages for the whole primary RC sample.
The values estimated from spectra of low-quality may suffer large systematics (e.g. Xiang et al. 2017a and Wu et al. 2019), since the KPCA-based multivariate linear regression approach is quite sensitive to the quality of the spectra (e.g. SNR).
As in Xiang et al. (2017a) and Wu et al. (2019), internal calibrations are performed for the derived masses and ages using duplicate observations.
To do so, a parameter $d_{\rm g}$ is defined.
This parameter is given by the maximal kernel value between the target spectrum and the spectra in the training set, and describes their similarities. 
The value of $d_{\rm g}$ can vary from 0 to 1, with unity representing exactly the same between a target spectrum and one in the training spectra.
Generally, there is a good relation between $d_{\rm g}$ and SNR, but $d_{\rm g}$ is found to be more efficient than SNR for internal calibration.
Small values of $d_{\rm g}$ appear due to little similarities between a target spectra and those in the training set.
The reason of the little similarities could be either the low spectral SNRs, or some unexpected features (e.g., residuary cosmic rays and/or sky lines) in the target spectra.

To calibrate masses and ages estimated with a small value of $d_{\rm g}$, duplicated observations with one yielding $d_{\rm g} \ge 0.8$ and another yielding  $d_{\rm g} < 0.8$ are adopted.
Parameters deduced from spectra of $d_{\rm g} \ge 0.8$ are assumed to be free of the systematics and thus are adopted as reference values for the calibration.
Those parameters derived from the latter observations that yield $d_{\rm g} < 0.80$ are divided into different bins of $d_{\rm g}$ with a bin size of 0.05 for $0.50 \le d_{\rm g} < 0.80$ and 0.075 for  $0.30 \le d_{\rm g} < 0.50$.
Results from spectra with $d_{\rm g} < 0.30$ are discarded considering the potential large errors.
A total of about 10,000 stars are excluded in this way.
For each $d_{\rm g}$ bin, a first-order polynomial is applied to calibrate masses and ages estimated from spectra with $d_{\rm g} < 0.80$  to those from spectra with $d_{\rm g} \ge 0.80$.
Stellar masses and ages of nearly 60,000 stars with $d_{\rm g} < 0.80$ are internally calibrated in this way.

\subsection{Validation of estimated masses and ages}
In this Section, we examine the accuracies of the masses and ages of primary RCs as determined above.
First, the internal uncertainties are evaluated using the duplicate observations.
Secondly, external checks are performed by member stars of well-studied open clusters (OCs).

%internal check
To asses the uncertainties of estimated masses and ages for primary RCs, results from duplicate observations of similar spectral SNRs (differed by less than 20 per cent) and collected in different nights, are used.
The relative mass and age estimate residuals (after divided by $\sqrt{2}$) as a function of mean spectral SNR are shown in Fig.\,9.
The dispersions of the residuals show significant variations with SNR over 15 (mass) and 45 (age) per cent for SNRs smaller than 30, and 5-8 (mass) and 10 (age) per cent for SNRs larger than 100.
To assign proper random errors to the estimated masses and ages, we fit the relative residuals by,
\begin{equation}
\sigma_{\rm r} = a + \frac{c}{{\rm SNR}^{b}}\text{,}
\end{equation}
where $\sigma_{\rm r}$ represents the random error.
The resulting fit coefficients for mass and age  estimates are presented in Table\,1. 
In addition to the random errors, method errors are also considered and the final errors are given by $\sqrt{\sigma_{\rm r}^2 + \sigma_{\rm m}^2}$.
Here $\sigma_{\rm m}$ denotes the method error and the errors are 13 and 27 per cent (from the training residuals, see Section\,4.1       ), respectively, for mass and age estimates.
Most of the stars in our primary RC sample have spectral SNRs higher than 30 and thus the uncertainties are dominated by $\sigma_{\rm m}$.

\begin{table*}
\centering
\caption{Fit coefficients of the $K_{\rm s}$ band absolute magnitude calibration}
\begin{tabular}{ccccc}
\hline
[Fe/H]& $a$ &$b$ &$c$&$d$\\
\hline
$> -0.10$&$-1.643$&$-5.261\times 10^{-3}$&$2.045 \times 10^{-3}$&$-5.387 \times 10^{-5}$\\
$-0.25$,$-0.10$&$-1.478$&$-7.120\times 10^{-2}$&$1.103 \times 10^{-2}$&$-4.370 \times 10^{-4}$\\
$-0.50$,$-0.25$&$-1.406$&$-1.202\times 10^{-1}$&$1.938 \times 10^{-2}$&$-8.098 \times 10^{-4}$\\
$< -0.50$&$-1.862$&$1.282\times 10^{-2}$&$9.397 \times 10^{-3}$&$-6.046 \times 10^{-4}$\\
\hline
\end{tabular}
\end{table*}

%external check
Member stars of a given open cluster (OC) are usually believed to form from a single giant molecular cloud and thus have similar ages, distances, kinematics and chemical compositions.
Therefore, they serve as good testbeds to check the accuracy of our age determinations for primary RCs, as well as for distance and other parameter estimates yielded by LSP3 (Xiang et al. 2017a). 
To do so, seven OCs of a wide age range (1--10\,Gyr) observed by the LAMOST surveys are selected.
Then member stars of those OCs are picked out from our primary RC sample.
As an example, the selection of member stars in the RC evolutionary stage for NGC\,2682 is illustrated in Fig.\,10.
First, we select candidates from our primary RC sample within four cluster diameters from the center of the OC.
The diameters and central positions of the OCs are adopted from the catalog of Dias et al. (2012).
Secondly, only stars located close to the red clump region of the appropriate cluster isochrone in the color magnitude diagram (CMD), i.e. the clump region, are selected.
Finally, we require the member stars must have proper motions similar to the cluster values given in the Gaia DR2, $|\mu_\alpha - \mu_\alpha^{\rm cluster}| < 3\sigma_{\mu_\alpha^{\rm cluster}}$ and $|\mu_\delta - \mu_\delta^{\rm cluster}| < 3\sigma_{\mu_\delta^{\rm cluster}}$.
%similar radial velocities as reported in the literature ($|v_{\rm los} -  v_{\rm los}^{\rm cluster}| \leq 10$\,km\,s$^{-1}$). 
In total, 22 member stars are selected for the seven OCs .
In Fig.\,11, the weighted mean ages (see Table\,2) given by our KPCA technique for the seven OCs are compared to the literature values and they generally agree with each other.

\section{New $M_{K_{\rm s}}$ calibration and distances of primary RCs}
\subsection{New $M_{K_{\rm s}}$ calibration of primary RCs}
To derive distances of the selected primary RCs, an accurate calibration of RC absolute magnitudes is required.
Based on the {\it Hipparcos} parallaxes and the open cluster member stars, absolute magnitudes of RCs in different bands (e.g. $VIJHK_{\rm s}$) have been calibrated previously by a number of studies (e.g. Paczy{\'n}ski \& Stanek 1998; Stanek \& Garnavich 1998;  Grocholski \& Sarajedini 2002; Salaris \& Girardi 2002; Groenewegen 2008; Laney et al. 2012; hereafter L12).
Generally, the absolute magnitudes in different bands are found to be constant and in the near-infrared bands, especially the $K_{\rm s}$ band, the absolute magnitudes are believed to be most stable least affected by the population effects. 
The $K_{\rm s}$ band is also less affected by the interstellar reddening compared to the visual bands (e.g. Salaris \& Girardi 2002;  L12).
Nevertheless, even for the near infrared bands, ignoring the population effects (i.e. metallicity and age) may introduce systematics on the level of, 5-10 per cent (see Fig.\,3 in Salaris \& Girardi 2002 and also Fig.\,6 in Girardi 2016). 

To further improve the calibration, we have collected a large sample of common stars of the Gaia DR2 and our RC sample to re-calibrate $K_{\rm s}$ absolute magnitude of RCs as a function, for the first time, of both metallicity and age.
The sample is constructed with the following cuts:
\begin{itemize}[leftmargin=*]

\item The stars must have a Galactic latitude $|b| \ge 10^{\circ}$ and a value of $E (B - V)$ less than 0.1\,mag, either from Schlegel, Finkbeiner \& Davis (1998; hereafter SFD98) for high latitudes ($|b| \ge 30^{\circ}$) or estimated with the ‘star pair’ technique (Yuan, Liu \& Xiang 2013) for stars close to the disk plane ($10 < |b| < 30^{\circ}$).
%, in order to minimise uncertainties due to reddening corrections;

\item The stars must have LAMOST spectra of SNRs higher than 60 and relative uncertainties of age, as determined above, less than 40\,per cent;

\item The photometric errors in the $K_{\rm s}$ band must smaller than 0.03\,mag;

\item The stars have distances estimated by Sch{\"o}nrich, McMillan and Eyer (2019; hereafter SME19), using the Gaia DR2 parallaxes (Lindegren et al. 2018). 
Note that in the treatment of SME19, the Gaia parallax uncertainties have been icreased by 0.043\,mas in quadrature and corrected for a parallax offset of 0.054\,mas; 

\item Following the suggestions of SME19, further cuts on the Gaia measurements have been applied to minimize the biases in the estimated distances: 1) Relative parallax uncertainties smaller than 10 per cent; 2) Parallax uncertainties smaller than 0.05\,mas; 3) Estimated distances greater than 80\,pc;  4) $G$ band magnitudes smaller than 14.0\,mag and $0.5 < G_{\rm BP} - G_{\rm RP} < 1.4$\,mag; 5) The number of measurements $n_{\rm} > 5$ and excess noise\,$< 1$; and 5) BP$-$RP excess flux factor $1.172 < E_{\rm BPRP} < 1.3$.

\end{itemize}
The first and last three cuts are to ensure precise determinations of $K_{\rm s}$ band absolute magnitudes.
The second one is to ensure low contamination from the secondary RCs and RGBs, as well as to ensure accuracies of the estimated metallicity and age.
Finally, a total of 13\,764 stars have been selected.
Their $K_{\rm s}$ band absolute magnitudes are derived using the distances estimated by SME19 from the Gaia DR2 parallaxes (Lindegren et al. 2018), and the $K_{\rm s}$ magnitudes from 2MASS (Skrutskie et al. 2006) after reddening corrections.
$K_{\rm s}$ band absolute magnitudes of primary RCs as a function of age are presented in Fig.\,12 and a significant trend is detected for primary RCs of ages greater than 5\,Gyr.
By further grouping the stars into different metallicity bins, we show the median values of $M_{K_{\rm s}}$ for the individual metallicity and age bins.
Generally, the trends of variations with metallicity and age are in excellent agreement with the theoretical predictions of Salaris \& Girardi (2002).
For stars with age greater than 4-5\,Gyr, the $K_{\rm s}$ band absolute magnitudes of primary RCs decrease with age by few hundredth maglitude per Gyr and increase with [Fe/H] also by few hundredth maglitude per dex\footnote{Note, for stars of [Fe/H]$> -0.10$, the dependence of $M_{K_{\rm s}}$ on [Fe/H] tends to be minor.}.
We note a similar trend of $M_{K_{\rm s}}$ with age is also found by Chen et al. (2017), based on $\sim$170 seismically identified RCs of solar metallicity in the {\it Kepler} field. 
For stars of ages younger than 5\,Gyr, $M_{K_{\rm s}}$ show weak dependence on both age and metallicity, with a median comparable to that  {\it Hipparcos} local sample (e.g. L12).
The typical standard deviation, after subtracting contribution from the photometric and astrometric uncertainties, is around 0.15\,mag (i.e. 7 per cent in distance).
To consider both the metallicity and age effects, we use a third-order polynomial to fit $M_{K_{\rm s}}$ as a function of age for different metallicity bins,
\begin{equation}
\begin{split}
M_{K_{\rm s}} = &a - b \tau + c\tau^2 + d\tau^3 \text{.}
\end{split}
\end{equation}
Here $\tau$ is stellar age. 
The coefficients for the different metallicity bins are given in Table\,3.

\subsection{Distances of primary RCs}
Using the newly calibrated relation, distances have been derived for all the selected primary RCs using the LSP3 values of [Fe/H], stellar ages estimated in Section\,4 and $K_{\rm s}$ band magnitudes from 2MASS (Skrutskie et al. 2006), after applying corrections for the reddening as given by SFD98 for stars of high latitudes ($|b| \ge 30^{\circ}$) or derived with the ‘star pair’ technique (Yuan, Liu \& Xiang 2013) for stars close to the disk plane ($|b| < 30^{\circ}$).

We further validate the distances estimated from our new calibration using the statistical method of Sch{\"o}nrich, Binney \& Asplund (2012; hereafter SBA12).
As shown in the upper panel of Fig.\,13, the potential distance bias estimated with SBA12 technique as a function of age for the different metallicity bins are all within a few (no more than 5)  per cent.
The results show that our new calibration is very accurate for primary RCs of all populations. 
In the bottom panel of Fig.\,13, we show the same results but plotted against the estimated distances.
Again, the zero-offsets of distances estimated with our new calibration are within 5 per cent for essentially all the distance bins.
For the distances derived assuming a constant value of $M_{K_{\rm s}}$ as calibrated by L12 using {\it Hipparcos} local sample, the resulted distance are biased less than 5 per cent for $d < 2$\,kpc but overestimated by  5-10 per cent for $d > 2$\,kpc.
The results are easily understandable. 
The local RCs ($d < 2$\,kpc) have ages and metallicities similar to those of the {\it Hipparcos} local sample adopted by L12, leading to minor distance biases.
For more distant RCs ($d > 2$\,kpc), their ages and metallicities may deviate significantly from those of the {\it Hipparcos} local sample, resulted in significant distances biases using the calibration of $M_{K_{\rm s}}$ by L12.
In addition, similar examinations are applied against $J - K_{\rm s}$ color and metallicity [Fe/H] of RC stars.
The results again show the systematic biases of the distance estimated by our new calibration are within 5 per cent for almost all the color and metallicity bins (see Fig.\,A1). 
We also examine the distances of primary RCs selected from the APOGEE DR14 (Bovy et al. 2014), using essentially the same technique described in the current work.
In their work, the distances have been derived using the PARSEC stellar isochrones, scaled to the local calibration of L12 by adding a constant offset (see Section\,3 of Bovy et al. 2014 for details).
For the local RCs ($d < 2$\,kpc), the distance biases are minor, on the level of a few per cent.
But to our surprise, the biases increase almost linearly with distance for more distant RCs of $d > 2$\,kpc and reach over 40 per cent at $d \sim 4.5$\,kpc.

\section{selection function of the sample}
To assess the selection function of our primary RC sample, we follow the procedure of Chen et al. (2018).
We calculate the selection function of the different fields (each field has an area around 3.36\,sq. deg., partitioned with {\it HEALPix}\footnote{https://healpix.sourceforge.io/};) by comparing the number of stars observed with LAMOST and with robust stellar parameter estimates  to that given by the completed 2MASS photometric catalog in the $J - K_{\rm s}$ versus $K_{\rm s}$ plane.
The final results are shown in Fig.\,14. 
One can see significant variations of target selection effects across the $J - K_{\rm s}$ versus $K_{\rm s}$ diagram.
Generally, the fainter the magnitude and the redder the color, the lower sampling rate.
The reasons for the trend are that we have used only blue-arm LAMOST spectra in deriving the stellar parameters and also applied a spectral SNR (at 4650\,\AA) cut ($\ge 20$) in selecting the primary RCs.
For more details about the target selection and the selection function of LAMOST data, we refer to Chen et al. (2018).

\section{Properties of The LAMOST RC sample}
\subsection{Spatial coverage}
Fig.\,15 shows the number density distributions of tour sample primary RCs in the $X$-$Y$ and $X$-$Z$ planes\footnote{Here $X$, $Y$ and $Z$ are axes of a Galactocentric, right-handed Cartesian system, with $X$ pointing in the direction opposite to the Sun, $Y$in the direction of Galactic rotation and $Z$ towards the North galactic Pole.}.
The sample covers a large volume of the Galactic disk, with $R$ (the projected Galactocentric distance in a Galactocentric cylindrical system) ranging from 4 to 16\,kpc, $Z$ from $-5$ to 5 kpc and $\phi$ (in the direction of Galactic rotation) from $-20$ to 50$^{\circ}$.
This large volume should certainly allow one to probe the structural, chemical and kinematic properties of both the Galactic thin and thick disks.

\subsection{Age distribution}
The distribution of mean stellar ages for $2.5^{\circ} \times 2.5^{\circ}$ projected patches on the sky in the Galactic coordinate system is presented in Fig.\,16.
As expected, a positive age gradient with increasing absolute Galactic latitude ($|b|$) is clearly seen.
Generally, the mean age is younger than 3\,Gyr for $|b| < 10^{\circ}$ and can be as old as 8\,Gyr at higher latitudes, say $|b| > 50^{\circ}$.  
Fig.\,17 shows the mean [$\alpha$/Fe] ratio and age of stars at different positions across the $R$-$Z$ plane of the Galactic disk.
Generally, both mean [$\alpha$/Fe] ratio and age show a negative gradient in the radial direction and a positive gradient in the vertical direction. 
Strong flaring of the young Galactic disk is clearly detected in both distributions, with younger populations (indicated either by the [$\alpha$/Fe] ratio or the stellar age) extended to higher heights of the Galactic plane as $R$ increases.

The radial and vertical age gradients of the Galactic disk have been reported by Casagrande et al. (2016) and Martig et al. (2016b), respectively. 
Using a large sample of stars with ages derived from the LAMOST spectra, Xiang et al. (2017c) and Wu et al. (2019) present similar age maps similar to those reported here. 
By careful considering the selection effects of the  primary RC sample (see Section\,6), the structure of the Galactic disk of different mono-abundance or mono-age populations can be studied in detail with the current sample (Yu et al. 2020).

\subsection{Age, [Fe/H] and [$\alpha$/Fe] relations}
Fig.\,18 plots the distributions of stellar number densities and mean stellar ages in the [Fe/H]-[$\alpha$/Fe] plane.
As expected, a bimodal distribution of [$\alpha$/Fe] is clearly seen.
By applying an empirical cut in the [Fe/H]-[$\alpha$/Fe] plane, one finds 15 per cent stars belonging to the high [$\alpha$/Fe] sequence, i.e. that associated with the so-called chemical thick disk. 
In Fig.\,18, stars of the high [$\alpha$/Fe] sequence typically have ages older than 9-10 Gyr. 
In contrast, stars of the  in low [$\alpha$/Fe] sequence have much younger ages.
Our results are in excellent agreement with the previous finding for the solar neighborhood based on high resolution spectroscopy (e.g. Fuhrmann 1998; Bensby et al. 2003; Haywood 2008, Haywood et al. 2013; Hayden et al. 2015).

The middle panels of Fig.\,18 show the distributions of stellar number densities and mean values of [$\alpha$/Fe] in the age-[Fe/H] plane.
Generally, young ($< 8$\,Gyr) populations show a wide [Fe/H] range while there is an obvious lack of metal-rich stars in the old ($\ge 8$\,Gyr) populations. 
Such an age-metallicity relation is consistent with the results found for stars in the solar neighborhood (e.g. Haywood et al. 2013; Bergemann et al. 2014) and also from other large samples of stars (e.g. Xiang et al. 2017c and Wu et al. 2019).
The lack of a tight age-metallicity relation for young populations could be explained by the effects of stellar radial migration (e.g. Sellwood \& Binney 2002; Sch{\"o}nrich \& Binney 2009).

Finally, the bottom panels of Fig.\,18 show the distributions of stellar number densities and mean values of [Fe/H] in the age-[$\alpha$/Fe] plane.
Two sequences  are clearly seen -- one of the young ($<$\,$8$-$9$\,Gyr) populations with low [$\alpha$/Fe] ratios and another of older ($\ge 8$\,Gyr) populations with high values of [$\alpha$/Fe].
The underlying physics leading to the two sequences is still under hot debate (e.g. Haywood et al. 2013) and our current sample should certainly help solve the issue.
In addition to the two sequences, an obvious excess of stars of age $\leq 6$\,Gyr and [$\alpha$/Fe]\,$\ge 0.15$ (the so-called young $\alpha$-enhanced stars; e.g. Martig et al. 2015; Chiappini et al. 2015) is detected in the age-[$\alpha$/Fe] plane.
This pattern is also seen in our training sample (see the bottom-left panel of Fig.\,6).
We will discuss the origin(s) of those stars in a separate paper (Sun et al. 2020, in preparation).

Before summarizing, it is worth mentioning that we have started a series of studies of the Galactic disk(s) using this primary RC sample, including: 1) mapping the 3D asymmetrical kinematics and detecting new substructures in the Galactic disk (Wang et al. 2019, 2020);  2) determining the structural properties of different mono-abundance disk populations (Yu et al. 2020); 3) detecting kinematic signatures of the Galactic warp (Li et al. 2020, in preparation); and 4) exploring the origin(s) of the so-called young $\alpha$-enhanced stars (Sun et al. 2020, in preparation).

\section{Summary}
Based on stellar atmospheric parameters deduced with the latest version of LSP3 for whole LMDR4, nearly 140,000 primary RC stars of spectral SNRs higher than 20 have been successfully singled out, based on their positions in the metallicity-dependent effective temperature--surface gravity and color--metallicity diagrams, supervised by a high-quality asteroseismology dataset.
Based on the various tests, a purity and a completeness of over 85 per cent have been achieved for this current sample of primary RCs.

Using the relations trained by a KPCA method with thousands of RC stars in the  LAMOST-$Kepler$ fields that have accurate asteroseismic mass measurements, stellar masses and ages are further determined from the LAMOST spectra for our primary RC sample stars.
Various tests show typical uncertainties of 15 and 30 per cent, respectively, for the estimated masses and ages.

Using over ten thousand primary RCs with accurate distance measurements from the Gaia DR2 parallaxes, the $K_{\rm s}$ band absolute magnitudes of primary RC are re-calibrated, by considering the effects of both metallicity and age, for the first time.
With this new calibration, very accurate distances are derived for the whole sample, with typical uncertainties of 5-10 per cent, even better than the Gaia measurements for stars beyond 3-4 kpc.

The sample covers a significant volume of the Galactic disk of $4 \leq R \leq 16$\,kpc, $|Z| \leq 5$\,kpc, and $-20 \leq \phi \leq 50^{\circ}$.
Stellar parameters, line-of-sight velocities and elemental abundances deduced from the LAMOST spectra and proper motions from the Gaia DR2 are also provided in the whole sample stars.
The sample is of vital importance to probe the structural, chemical and kinematic properties of the Galactic disk(s) and is available online.

 \section*{Acknowledgements} 
The Guoshoujing Telescope (the Large Sky Area Multi-Object Fiber Spectroscopic Telescope, LAMOST) is a National Major Scientific Project built by the Chinese Academy of Sciences. Funding for the project has been provided by the National Development and Reform Commission. LAMOST is operated and managed by the National Astronomical Observatories, Chinese Academy of Sciences.

This work has made use of data from the European Space Agency (ESA) mission Gaia (https://www.cosmos.esa.int/gaia), processed by the Gaia Data Processing and Analysis Consortium (DPAC, https://www.cosmos.esa.int/web/gaia/dpac/consortium).

 It is a pleasure to thank J. Binney for a thorough read of the manuscript and helpful comments.

This work is supported by National Natural Science Foundation of China grants 11903027, 11973001, 11833006, 11811530289, U1731108, and U1531244, and  National Key R\&D Program of China No. 2019YFA0405503. 
Y.H. is supported by the Yunnan University grant C176220100007. 
R.S. is supported by a Royal Society University Research Fellowship.
H.F.W. is supported by the LAMOST Fellow project, National Key Basic Research Program of China via SQ2019YFA040021 and funded by China Postdoctoral Science Foundation via grant 2019M653504 and Yunnan province postdoctoral Directed culture Foundation. 
H.F.W. is also supported by the Cultivation Project for LAMOST Scientific Payoff and Research Achievement of CAMS-CAS. 

\appendix
\section{Distance biases versus color and metallicity of RCs}
The biases of the distances estimated by our new calibration (see Section\,5.1) against the color $J - K_{\rm s}$ and metallicity [Fe/H] of RCs are shown in Fig.\,A1.
\begin{figure}
\begin{center}
\includegraphics[scale=0.325,angle=0]{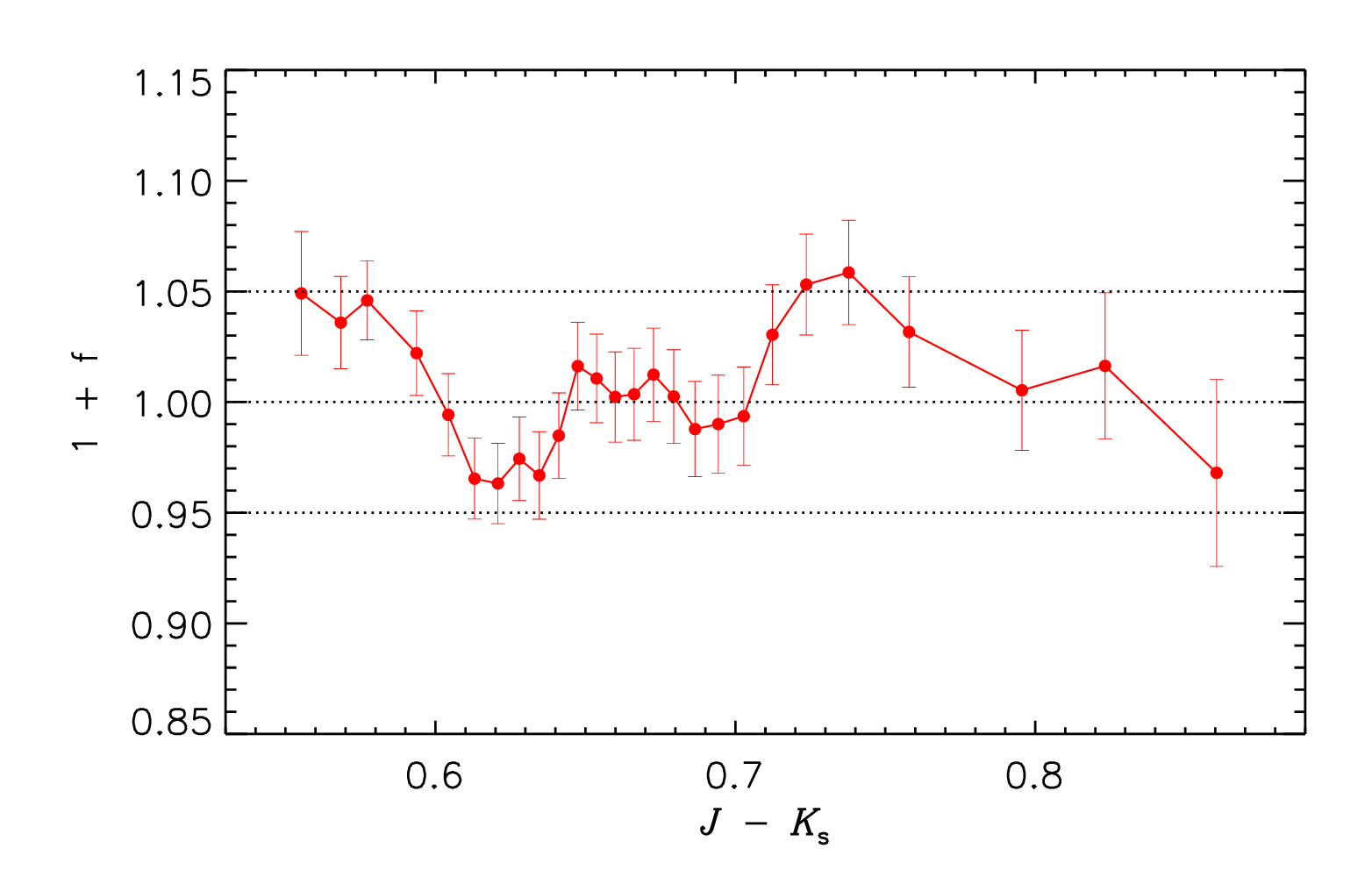}
\includegraphics[scale=0.325,angle=0]{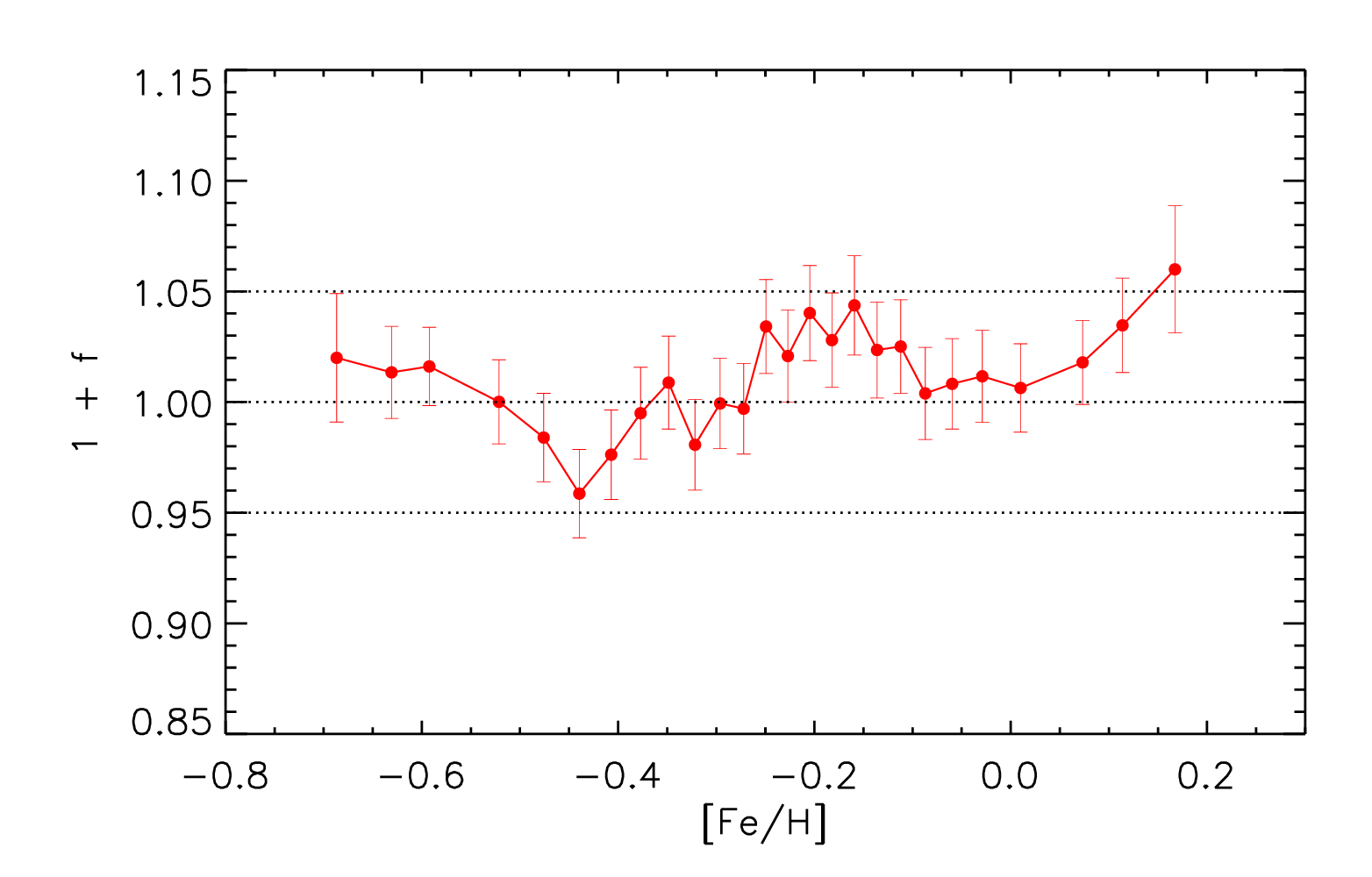}
\caption{{\it Upper panel:} relative error versus $J - K_{\rm s}$ for the LAMOST primary RCs of SNR\,$> 60$. 
The binning scheme is the same as for Fig.\,13.
{\it Lower panel:} relative error versus [Fe/H] for he LAMOST primary RCs of SNR\,$> 60$.
The binning scheme is again the same as for Fig.\,13.}
\end{center}
\end{figure}

%%% that appears after it.

\end{document}